\documentclass[%
 aip,
 amsmath,amssymb,
 reprint,%
]{revtex4-1}

\usepackage{graphicx}% Include figure files
\usepackage{xcolor}
\usepackage{dcolumn}% Align table columns on decimal point
\usepackage{bm}% bold math
\usepackage[utf8]{inputenc}
\usepackage[T1]{fontenc}
\usepackage{mathptmx}
\usepackage{amsmath}

\usepackage{ulem}
\begin{document}

\preprint{AIP/123-QED}
\title{Compact atomic descriptors enable accurate predictions via linear models}
\author{Claudio Zeni}
\affiliation{Physics Area, International School for Advanced Studies,  Trieste, IT}
  \email{czeni@sissa.it}
\author{Kevin Rossi}%
\affiliation{Laboratory of Nanochemistry, Institute of Chemistry and Chemical Engineering, Ecole Polytechnique Fédérale de Lausanne, Lausanne, CH}%
\author{Aldo Glielmo}
\affiliation{Physics Area, International School for Advanced Studies,  Trieste, IT}
\author{Stefano de Gironcoli}
\affiliation{Physics Area, International School for Advanced Studies,  Trieste, IT}

\date{\today}

\begin{abstract}
We probe the accuracy of linear ridge regression employing a 3-body local density representation deriving from the atomic cluster expansion.
We benchmark the accuracy of this framework in the prediction of formation energies in molecules, and forces-energies in solids.
We find that such a simple regression framework performs on par with state-of-the-art machine learning methods, which are, in most cases, more complex and more computationally demanding.
Subsequently, we look for ways to sparsify the descriptor and further improve the computational efficiency of the method. 
To this aim, we use both PCA and LASSO regression for energy fitting on six single-elements datasets.
Both methods highlight the possibility to construct a descriptor that is 4 times smaller than the original, with an equal or even improved accuracy.
Furthermore, we find that the reduced descriptors share a sizable fraction of their features across the six independent datasets, hinting at the possibility to design material-agnostic, optimally compressed, and accurate, descriptors.

% Please include a maximum of seven keywords
\end{abstract}

\maketitle

\section{\label{sec:intro}Introduction}

The advent of machine learning (ML) methods in atomistic simulation and modelling is benefiting a wide array of disciplines, e.g. it has become an important tool in the field of structure-property predictions. \cite{Smith2017,Artrith2019,Schleder2019}
As paradigmatic examples, accurate data-driven prediction of properties from structures have been developed for NMR shieldings in molecules and molecular crystals,\cite{Paruzzo2018,Gupta2021} polarization of small to medium molecular systems \cite{wilkins2019}, solvation and efficacy of drugs \cite{Rauer2020,Wu2018,Axelrod2020}, and activity of homogeneous and heterogeneous catalysts.\cite{Jager2018,Meyer2018,Gu2020}
By the same token, and in relationship to the development of force fields (FFs), representative achievements may be found in the simulation of reactions in solutions explicitly accounting for the solvent, \cite{Rossi2020,Yang2020} the assessment of the stability of multi-phase materials relevant, e.g., to storage-and-conversion \cite{Eckhoff2020,vandermause2020fly,Zeni2019}, electronic  devices \cite{Sosso2012,Zeni2018,Deringer2021}, geology,\cite{Zhang2020}, and to the realistic modelling of complex systems in soft matter and biophysics \cite{Lahey2020,Wang2019,Scherer2020}.

An open issue of particular importance in data driven approaches for atomistic systems lies in the choice of the representation of the atomistic system itself.
As a witness of the relevance of this problem, a multitude of atomic environment descriptors have been proposed in the last 15 years.
\cite{Behler:2007fe, Bartok2010, Rupp2012, Thompson:2015dw, Shapeev2016, Glielmo2018, Rossi2020c} 
Among the most successful representations in the field we find local density representations.
In a nutshell, these representations hinge on a construction where atom-centered distributions are represented in a vector form using a many-body expansion. \cite{Behler:2007fe, Bartok2010, Rupp2012, Thompson:2015dw, Shapeev2016, Glielmo2018}
Recently, a general formulation of such local density representation, named ''atomic cluster expansion`` (ACE), has been proposed by Drautz. \cite{Drautz2019, Bachmayr2020, Drautz2020}
The ACE representation is symmetric w.r.t. rotation, translation, and permutation of identical atoms.
It is furthermore differentiable w.r.t. atomic coordinates and complete, that is, in its generalized formulation it leads to a descriptor body-order which is iteratively expanded up to the desired one, hence satisfying the uniqueness principle.\\

In this work, we discuss the performance on two benchmark datasets of 3-body representations following from the ACE representation, used in conjunction with a ridge regression fitting procedure.
In Sec.~\ref{sec:Method}, we present the ACE descriptor \cite{Drautz2019}, the scaled (SC) and non-scaled (NSC) versions of the Chebyshev radial basis functions, and the simplified spherical Bessel (SSB) radial basis functions, first introduced in Kocer et al. for local atomic environments \cite{Kocer2019}.
We then introduce the regression algorithm used to predict atomic forces and total energies throughout the manuscript in Subsec.~\ref{subsec:Regression}.
The proposed descriptor-regression framework resembles the Spectral Neighbor Analysis (SNAP) Potentials first introduced by Ref.~\onlinecite{Thompson:2015dw}, but relies on power spectrum coefficients rather than bispectrum coefficients, making it a 3-body potential in the sense of Ref.~\onlinecite{Glielmo2019} rather than a 4-body potential (4+7-body in the case of quadratic SNAP).
Our regression framework is then benchmarked on two publicly available datasets in Sec.~\ref{sec:Benchmarks}.
Firstly, in Subsec.~\ref{subsec:QM9}, we consider the QM9 dataset, which contains atomic structures and properties, such as formation energy, of 134k small molecules. \cite{Rupp2012, Blum2009} 
We show that a simple ridge regression framework yields predictions for molecular systems that display an accuracy comparable to the one of more complex, and computationally demanding, regression methods.
Similarly to other local density representation methods, we observe a trade-off between computational cost and accuracy, where accurate enough predictions are found only for a sufficiently large dimension of the descriptor; this verifies regardless of whether we employ SC, NSC, or SSB polynomials as the set of radial basis functions in the descriptor.
Nevertheless, we observe that SSB functions enable more accurate predictions than the other two radial basis functions when a low number of radial basis functions is employed.
Secondly, in Subsec.~\ref{subsec:Zuo}, we look at the fitting of a FF for six single-element crystalline systems utilising the dataset of forces and energies introduced by Ref.~\onlinecite{zuo2020performance}.
We find again that the proposed learning framework can perform on par with other state-of-the-art approaches, and that its accuracy depends on the dimension of the representation.
Interestingly, we find that employing SSB radial basis functions is often optimal for compact representations.
In Sec.~\ref{sec:Dimension}, we discuss methods to reduce the dimension of descriptors employed to fit energies in the example case of the database containing six single-element periodic systems \cite{zuo2020performance}.
We find that through both principal component analysis (PCA) dimensionality reduction, and least absolute shrinkage operator (LASSO) regression feature selection, we are able to match, and sometimes outperform, the accuracy obtained when using the full descriptor, while reducing its dimension by a factor of $\sim$~4.
The features selected by PCA and LASSO across the six single-element datasets are furthermore partially redundant, revealing an underlying material-agnostic structure to the relevant directions in the data space.
This insight could guide the informed design of optimally compact, and computationally efficient, local atomic environment descriptors.
Finally, the conclusions summarise the results and offer an outlook for future research aimed at improving the algorithm proposed in this manuscript.
\section{\label{sec:Method} Methods}
\subsection{\label{subsec:Representation}Atomic Environment Representation}
To construct the local atomic environment descriptor $\mathbf{q}(\rho)$ used throughout this manuscript, we first define the local atomic density $\rho(\mathbf{r})$, through standard procedure, as a sum of Dirac delta functions $\delta(\mathbf{r}_{ji} - \mathbf{r})$ centered on each atom surrounding a central atom $i$ within a cutoff $r_{c}$:
\begin{equation}
\rho_i(\mathbf{r}) = \sum_{j| r_{ji} \leq r_{c}} \delta(\mathbf{r}_{ji} - \mathbf{r}),
\label{eq:local_atomic_environment}
\end{equation}
where $\mathbf{r}_{ji}$ indicates the vector $(\mathbf{r}_j - \mathbf{r}_i)$, and $r_{ji}$ the magnitude of $\mathbf{r}_{ji}$.
The local atomic environment representation in Eq.~\ref{eq:local_atomic_environment} is already invariant to permutations of identical atoms and translations, but not to rotation; it is moreover not trivial to transform such representation into a finite-size descriptor.
To overcome these problems, the local atomic density is first approximated via a truncated expansion in spherical harmonics and radial basis functions:
\begin{equation}
\rho_i(\mathbf{r}) \sim \sum_{j \in \rho_i} \sum_{n=0}^{n_{MAX}} \sum_{l=0}^{l_{MAX}} \sum_{m=-l}^{l} c_{nlm}^j g_{n}(r_{ji}) Y_{lm}(\hat{\mathbf{r}}_{ji}),
\label{eq:spherical_harmonics_expansion}
\end{equation}
where $\hat{\mathbf{r}}_{ji}$ is the unit vector of $\mathbf{r}_{ji}$, $g_{n}$ are elements of a set of $n_{MAX}$ radial basis functions, $Y_{lm}$ are elements of a set of spherical harmonics, $n_{MAX}$ indicates the truncation limit for the radial basis set, and $l_{MAX}$ the truncation limit for the angular basis set.
We note that the elements $g_n$ should also depend on the angular expansion coefficient $l$.
We here remove the coupling between angular and radial parts following the approach of Ref.~\onlinecite{Kocer2019}, as it was shown that such simplification significantly reduces the complexity of evaluating $g(r_{ij})$ without noticeable decreases in the prediction accuracy \cite{Kocer2019, Kocer2020}.
In principle, one could use the array of coefficients $C_{nlm} = \sum_{j \in \rho_i} c_{nlm}^j$ as a descriptor, but it would not be invariant to rotations of the local atomic environment.
To solve this issue, products of $N$ coefficients $c_{nlm}^j$ that correspond
to a reducible representation of the identity of the rotation group are taken.
The resulting descriptors are of order $(N+1)$, i.e. can encode the interaction of up to $N+1$ atoms at once. \cite{Drautz2019, Glielmo2018}
One advantage of ACE descriptors is given by the linear scaling of the computational cost for their evaluation in the number of atoms $M$ in the neighborhood of $i$, for any order $N$.
This is not the case, e.g.,  for explicit $N$-body descriptors, where summations over groups of $N$ neighbors have to be considered, therefore causing the computational cost of their evaluation to scale as $\mathcal{O}(M^{N-1})$, i.e., more than linearly in $M$ whenever $N>2$. 
The linear scaling in $M$ of ACE descriptors therefore enables a large computational speed-up when compared to explicit $N$-body descriptors, especially for densely-packed systems.
In this manuscript, we employ 3-body descriptors,
where components $q_{n_1, n_2, l}$ of $\mathbf{q}$ are computed as:
\begin{equation}
q_{n_1, n_2, l}(\rho_i) = \sum_{j \in \rho_i}  \sum_{k \in \rho_i} \sum_{m=-l}^{m=+l} (-1)^m c^j_{n_1lm} c^k_{n_2l-m}.
\label{eq:B2_descriptor}
\end{equation}
The representation in Eq.~\ref{eq:B2_descriptor} is expected to strike a good balance between descriptiveness and efficiency \cite{Zeni2018, Glielmo2017}, as the computational complexity of evaluating the 3-body descriptor is $\mathcal{O}(M \cdot (n_{MAX} \cdot l_{MAX}^2 + n_{MAX}^2 \cdot l_{MAX}))$.
Nevertheless, we expect that the inclusion of 4- and higher-body descriptors, following the procedure in Ref.~\onlinecite{Drautz2019}, will enable to reach even higher accuracies, albeit at an increased computational cost.
The equations reported so far hold for single-element systems.
If $S>1$ atomic species are present, we employ $S^2$ independent descriptors $\mathbf{q}_{a, b}(\rho_i)$, where $a$ refers to the type of the central atom $i$, and only surrounding atoms of type $b$ contribute to the value of $\mathbf{q}_{a, b}(\rho_i)$.\\

In the first manuscript introducing the so-called  Atomic Cluster Expansion, Drautz proposes an ensemble of SC polynomials as the orthonormal radial basis set $g_n(r)$ \cite{Drautz2019}, which had been also previously used to expand 2- and 3-body correlation functions and chart structure-to-property mappings in Ref.~\onlinecite{Artrith2017}.
We report here the SC radial basis set as defined in Ref.~\onlinecite{Drautz2019}:
\begin{equation}
  \begin{gathered}
g_0(x) = 1,\\
g_1(x) = \dfrac{1}{2} \left[1+\text{cos}(\pi r/r_c\right)],\\
g_n(x) = \dfrac{1}{2} \left[1 - T_{n-1}(x)\right]  \dfrac{1}{2} \left[1+\text{cos}(\pi r/r_c)\right] ,
  \end{gathered}
\label{eq:Chebyshev_pt1}
\end{equation}
where the Chebyshev polynomials of the first kind $T_n(x)$ are defined recursively as:
\begin{equation}
  \begin{gathered}
T_0(x) = 1,\\
T_1(x) = x,\\
T_{n+1}(x) = 2 x T_n(x) - T_{n-1}(x),
  \end{gathered}
\label{eq:Chebyshev_pt2}
\end{equation}
and the scaled distance function is:
\begin{equation}
x = 1 - 2 \left( \dfrac{e^{-\lambda(r/r_{c}-1)}}{e^{\lambda} -1}  \right),
\label{eq:scaled_distance_Chebyshev}
\end{equation}
where $\lambda$ is a coefficient, set to 5 as in Ref.~\onlinecite{Drautz2019}.

Beside the SC radial basis, we look at possible changes in performance originating from the use of different radial basis in the ACE expansion: a set of SSB functions of the first kind, introduced in Ref.~\onlinecite{Kocer2019}, and a NSC radial basis set.
The NSC radial basis set is defined by Eq.~\ref{eq:Chebyshev_pt1} where $x$ is $x = 2r/r_c - 1$.
The SSB functions of the first kind basis set $g_n(r_{ij})$, introduced in Ref.~\onlinecite{Kocer2019}, is defined recursively as:
\begin{equation}
  \begin{gathered}
g_n(r) = \dfrac{1}{\sqrt{d_n}} \left(f_n(r) + \sqrt{\dfrac{e_n}{d_{n-1}}} g_{n-1}(r)   \right), \\
d_n = 1- \dfrac{e_n}{d_{n-1}}, \\
e_n = \dfrac{n^2 (n+2)^2}{4 (n+1)^4 +1},
  \end{gathered}
\label{eq:bessel}
\end{equation}
where $d_0 = 1$,  $d_1 = 1$, $g_0(r) = 1$, $g_1(r) = f_0(r)$, and:
\begin{multline}
f_n(r) = (-1)^n \dfrac{\sqrt{2} \pi}{r_c^{3/2}} \dfrac{(n+1)(n+2)}{\sqrt{(n+1)^2+(n+2)^2}} \cdot  \\
\left[\text{sinc}\left(r\dfrac{(n+1)\pi}{r_c} \right) + \text{sinc}\left(r\dfrac{(n+2)\pi}{r_{c}} \right) \right].
\label{eq:bessel_part2}
\end{multline}
\begin{figure}[t!]
    \centering
    \includegraphics[width=7.cm]{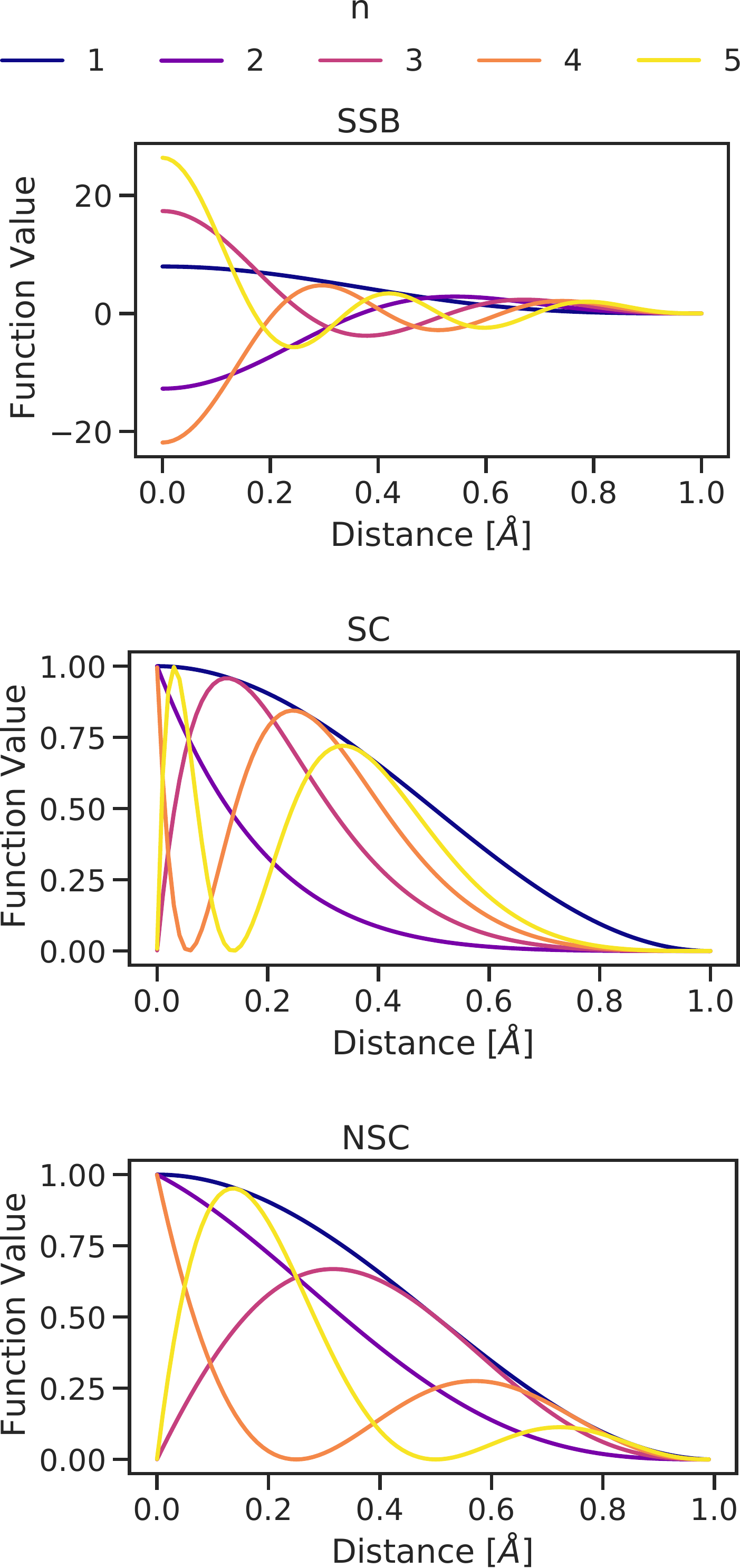}
    \caption{Visualisation of the terms $g_1$ to $g_5$ of (top to bottom) the SSB, SC, and NSC} radial basis function sets.
    The cutoff radius was set to 1 $\text{\AA}$ for the three plots.
    \label{fig:radial_basis}
\end{figure}
In Fig.~\ref{fig:radial_basis} we display the terms $g_1$ to $g_5$ for the three radial basis function sets, where $r_c$ was set to 1.\\

Additionally, independently of the choice of the radial basis set, we have found that appending the element-wise squared descriptor $\mathbf{q}^2$ to $\mathbf{q}$ yields a sizable increase in prediction accuracy, with negligible computational cost.
The further inclusion of $\mathbf{q}^3$, or higher-order powers of $\mathbf{q}$, element-wise square-root, or sigmoid, of the original descriptor $\mathbf{q}$, did not appear to yield any significant accuracy increase.

\subsection{\label{subsec:Regression}Regression Algorithm}
To carry out the supervised learning task we adopt ridge regression (RR), one of the simplest and most computationally efficient fitting algorithms.
RR recasts the learning problem into the following closed formula:
\begin{equation}
    \mathbf{Y} = \mathbf{Q} ~ \mathbf{W}  + \mathbf{\epsilon},
\label{eq:ridge_form}
\end{equation}
where $\mathbf{Y}$ is the matrix of dependent variables , $\mathbf{Q}$ is the matrix of explanatory variables, $\mathbf{W}$ is the parameter matrix that weights $\mathbf{Q}$, and $\mathbf{\epsilon}$ is a vector of error terms which accounts for possible hidden variables influencing $\mathbf{Y}$ that are not contained in $\mathbf{Q}$. 
Given a training set $\mathcal{D} = \{\mathbf{Y}_i, \mathbf{Q}_i \}~ i = 1, \dots, D$, we obtain the weights $\mathbf{W}$ analytically as:
\begin{equation}
\mathbf{W} = (\mathbf{Q}^T\mathbf{Q} + \gamma~ \mathbb{I})^{-1}\mathbf{Q}^T\mathbf{Y},
\label{eq:ridge_fitting}
\end{equation}
where $\gamma$ is the ridge parameter.
When both forces and energies are used to train the algorithm, $\mathbf{Y}$ is a 2D matrix with elements $\mathbf{Y}_d$ pertaining to structure $d$ containing $S$ atoms:
\begin{equation}
\mathbf{Y}_d = \left[ E_d, f^x_1, f^y_1, f^z_1, \dots, f^x_S, f^y_S, f^z_S \right], 
\label{eq:force_energy_fitting}
\end{equation}
where $f^c_s$ indicates the $c$-component of the force vector acting on atom $s$ of structure $d$.
Similarly, the matrix of explanatory variables $\mathbf{Q}$ becomes a 3D tensor with elements $\mathbf{Q}_d$ pertaining to structure $d$:
\begin{equation}
\mathbf{Q}_d = \left[ \mathbf{q}_d, -\dfrac{\partial \mathbf{q}_d}{\partial x_1}, -\dfrac{\partial \mathbf{q}_d}{\partial y_1}, -\dfrac{\partial \mathbf{q}_d}{\partial z_1}, \dots, -\dfrac{\partial \mathbf{q}_d}{\partial x_S}, -\dfrac{\partial \mathbf{q}_d}{\partial y_S}, -\dfrac{\partial \mathbf{q}_d}{\partial z_S} \right],
\label{eq:force_energy_descriptor}
\end{equation}
where $\mathbf{q}_d$ is defined as the sum over all atoms $i$ in structure $d$ of the local atomic environment descriptor $\mathbf{q}(\rho_i)$.
Whenever only energies are used to fit the algorithm, such as in the case of the QM9 dataset, both the elements of the matrix of explanatory variables and the elements of the matrix of dependant variables simplify to, respectively, $\mathbf{Q}_d = \mathbf{q}_d$ , and $\mathbf{Y}_d = E_d$. \\

Two main advantages arise from the choice of employing RR over more complex learning algorithms such as artificial neural networks (ANNs) or Gaussian Process (GP) regression.
Firstly, RR has a lower computational cost than ANNs or GP regression, since once the descriptor $\mathbf{Q}$ has been calculated, the prediction of $\mathbf{Y}$ requires a single matrix product.
Secondly, RR models, similarly to GPs, can be trained in closed-form and therefore without the need of slow gradient descent algorithms, which also introduce additional hyper-parameters that require careful tuning.

\section{\label{sec:Benchmarks}Results}
\subsection{\label{subsec:QM9}Energy prediction in the QM9 Dataset}
\begin{figure}[t!]
    \centering
    \includegraphics[width=7.5 cm]{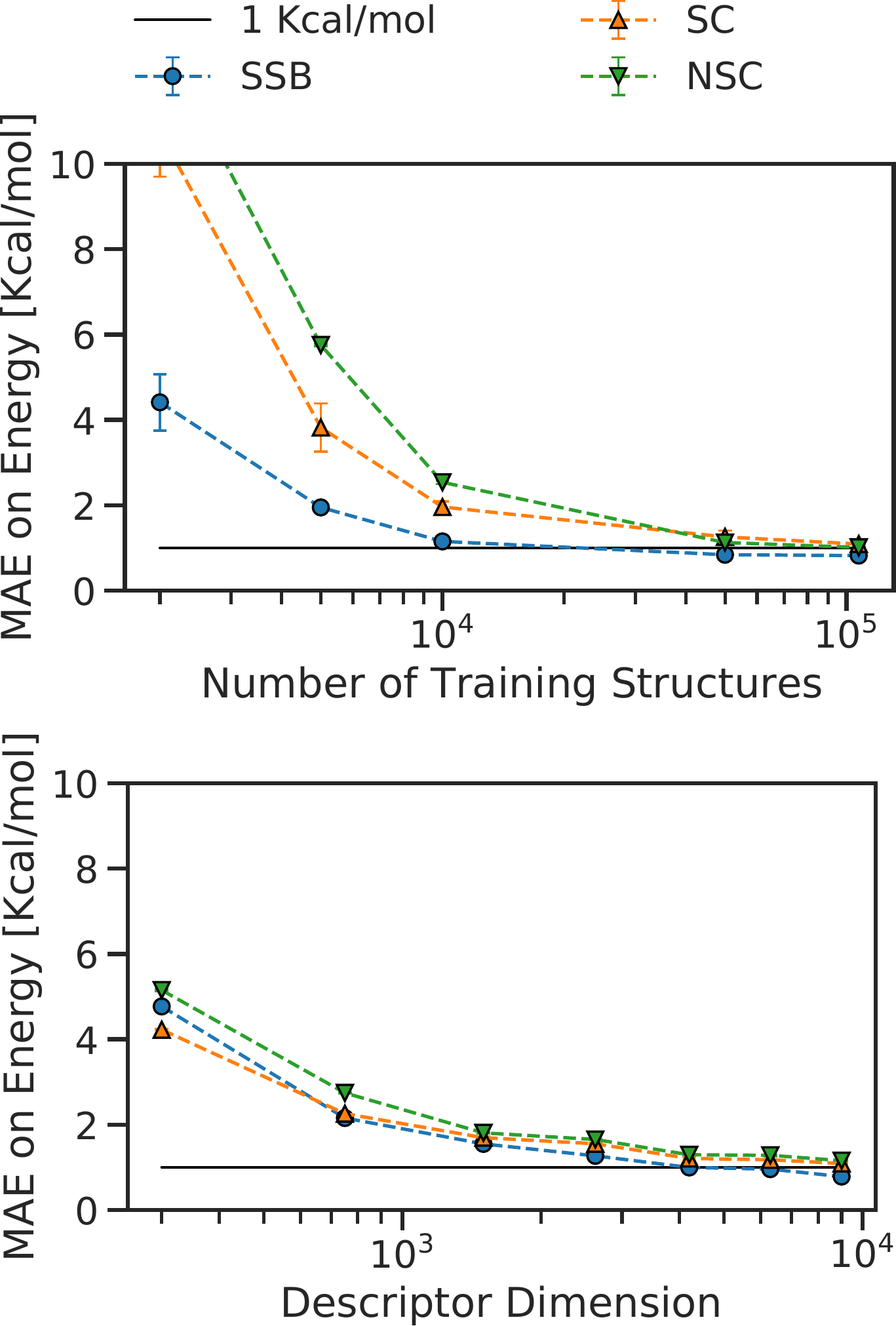}
    \caption{Validation MAE on total formation energy for the QM9 dataset incurred by potentials, as a function of the number of training structures used with $n_{MAX}$ = 8 and $l_{MAX}$ = 10 (top panel), and as a function of the number of features in the representation, with $l_{MAX}$ = $n_{MAX}$ + 2 and $n_{MAX}$ = $2, \dots, 8$ and using 107800 training structures (bottom panel).
    The standard deviation of each measure across three independent runs is displayed, where the validation and training set were randomly selected, and the size of the validation set was kept fixed at 26777 (20\% of the total dataset).
    The black dashed line indicates a MAE of 1 Kcal/mol, typically indicated as a target for chemical accuracy in structure energy prediction.
    }
    \label{fig:QM9}
\end{figure}

As a first benchmark, we look at one of the most widely studied datasets in our community, the QM9 dataset, \cite{Rupp2012, Blum2009, Ramakrishnan_2014} and aim to predict the formation energy for each molecule in the database.
The QM9 encompasses a relatively large number (133885) of molecules with a total of up to 5 chemical species, with each molecule containing up to 9 heavy atoms of C, N, O, or S, and any number of H atoms.
In the top panel of Fig.~\ref{fig:QM9}, we report the mean absolute error (MAE) incurred on total energy predictions by three sets of RR models employing different radial basis functions, for a fixed set of descriptor hyperparameters: $r_{C}$ = 4.5, $n_{MAX} = 8$, $l_{MAX} = 10$ and, therefore, a fixed descriptor's dimensionality.
A black dashed line in the plot indicates the target of 1 kcal/mol, often referred to as the target chemical accuracy for the the prediction of formation energies for molecules.
Among the three radial basis expansions under scrutiny, the SSB basis set performs best at any point on the training curve, and it reaches chemical accuracy, even when fewer than the maximum number of training structures (107800, 80\% of the total dataset) are used. 
We hypothesize that this result is a consequence of the higher spatial resolution of Bessel polynomials w.r.t. (scaled and non-scaled) Chebyshev polynomials for low values of $n_{MAX}$, as discussed in the Supplementary Material (SM), Section A.
The convergence error for the RR model here presented, employing SSB radial functions with $n_{MAX} = 8$, $l_{MAX} = 10$, is 0.78 $\pm$ 0.02 Kcal/mol.
Other methods are able to incur lower MAEs, such as neural networks, reaching 0.14 Kcal/mol in Ref.~\onlinecite{Unke2018}, or Gaussian process regression, reaching 0.14 Kcal/mol in Ref.~\onlinecite{Willatt2019}.
The lower error incurred by the aforementioned methods is however paralleled by a much higher computational cost and complexity.
Furthermore, we are not aware of any linear method which has been successfully used for the prediction of formation energies for the QM9 dataset, i.e. displaying a MAE on par with chemical accuracy.\\

As stated in Sec.~\ref{subsec:Representation}, the computational cost of the atomic cluster expansion descriptor strongly depends on the choice of $n_{MAX}$ and $l_{MAX}$, which in turn affects the descriptor's dimension.
For this reason, the investigation of the validation error as a function of the descriptor's dimension reveals the accuracy/cost trade-off of the algorithm, and can lead to increased efficiency.
The bottom panel of Fig.~\ref{fig:QM9} displays the validation MAE incurred by RR trained on 107800 structures for the QM9 dataset as a function of the descriptor's dimension when employing SSB, SC, and NSC radial basis functions.
In this instance too, RR FFs employing the SSB basis set reach chemical accuracy (1 Kcal/mol) at smaller descriptor dimensions than the other two basis sets, and more specifically at $n_{MAX} = 6$, $l_{MAX} = 8$.
This paradigmatic example shows that the choice of the most efficient basis may be key when developing surrogate models for databases which encompass a large number of data and of chemical species.
For an analysis of the impact on prediction accuracy of the balance between $n_{MAX}$ and $l_{MAX}$, the interested reader is directed to the SM, Section 2.

\subsection{\label{subsec:Zuo}Force and Energy Prediction in Materials}
In the previous paragraph, we benchmarked the accuracy of our method while fitting on formation energies only. 
While energy prediction is of great importance, e.g. for structure search methods, it is often the case that {\it both} forces and energies are required, e.g. when running molecular dynamics (MD) simulations using a ML FF.
In this second example, we consider the database containing forces and energies for six single-element periodic systems, first introduced in Ref.~\onlinecite{zuo2020performance}.
The dataset contains perfect and deformed crystalline structures for two group IV semiconductors (Si and Ge), two body-centred-cubic (BCC) metals (Li and Mo), and two face-centred-cubic (FCC) metals (Ni and Cu); for additional details on the methods used to generate the data, the interested reader is referred to Ref.~\onlinecite{zuo2020performance}.
We thus asses the accuracy of our framework hinging on RR fitting and a 3-body ACE representation to produce a FF given each of the six single-element datasets.
Tables \ref{tab:zuo_forces} and \ref{tab:zuo_energy} report the performance of the proposed ML framework, using a SSB radial basis employing $n_{MAX}$=8 and $l_{MAX}$=10, and using the same system dependent cut-offs $r_{c}$ used in Ref.~\onlinecite{zuo2020performance} ( Mo = 5.2 {\AA}, Si = 4.7 {\AA}, Ge = 5.1 {\AA}, Cu = 3.9 {\AA}, Ni = 4.0 {\AA}, Li = 5.1 {\AA}).
For reference, we also report the results from Ref.~\onlinecite{zuo2020performance}, which benchmarked other widespread state-of-the-art ML frameworks.
Notwithstanding the simplicity and computational efficiency inherent to a linear fit, the proposed approach displays performances comparable to the most accurate methods discussed in the literature.

\begin{table}[h!]
    \centering
    \begin{tabular}{c|c|c|c|c|c|c}
    Material & Our Method & GAP$^*$ & MTP$^*$ & NN-BP$^*$ & SNAP$^*$ & qSNAP$^*$ \\
    \hline
Ni  &	0.03 &	0.04 &	0.03 &	0.07 &	0.08 &	0.07 \\
Cu  &	0.02 &	0.02 &	0.01 &	0.06 &	0.08 &	0.05 \\
Li  &	0.01 &	0.01 &	0.01 &	0.06 &	0.04 &	0.04 \\
Mo  &	0.16 &	0.16 &	0.15 &	0.20 &	0.37 &	0.33 \\
Si  &	0.13 &	0.12 &	0.09 &	0.17 &	0.34 &	0.29 \\
Ge  &	0.09 &	0.08 &	0.07 &	0.12 &	0.29 &	0.20 \\
    \end{tabular}
    \caption{Minimum RMSE on forces (eV/$\text{\AA}$) incurred by our 3-body RR potential employing SSB polynomials as radial basis functions, for the six single element datasets from Ref.~\onlinecite{zuo2020performance}.
    The symbol $^*$ indicates results from Ref.~\onlinecite{zuo2020performance}, which are included for comparison.}
    \label{tab:zuo_forces}
\end{table}
\begin{table}[h!]
    \centering
    \begin{tabular}{c|c|c|c|c|c|c}
    Material & Our Method & GAP$^*$ & MTP$^*$ & NN-BP$^*$ & SNAP$^*$ & qSNAP$^*$ \\
    \hline
Ni  &	1.74 &	0.62 &	0.74 &	2.25 &	1.17 &	1.04 \\
Cu  &	1.19 &	0.56 &	0.52 &	1.68 &	0.87 &	1.16 \\
Li  &	1.23 &	0.63 &	0.76 &	0.98 &	1.31 &	0.85 \\
Mo  &	4.00 &	3.55 &	3.89 &	5.67 &	9.06 &	3.96 \\
Si  &	5.16 &	4.18 &	3.02 &	9.95 &	8.06 &	6.28 \\
Ge  &	11.62 &	4.47 &	3.68 &	10.95 &	10.96 &	10.55 \\
    \end{tabular}
    \caption{Minimum RMSE on energies (meV/atom) incurred by our 3-body linear potential employing SSB polynomials as radial basis functions, for the six single element datasets from Ref. \cite{zuo2020performance}.
    The symbol $^*$ indicates results from Ref.  \cite{zuo2020performance}, which are included for comparison.}
    \label{tab:zuo_energy}
\end{table}
\begin{figure}
\includegraphics[width=8.5cm]{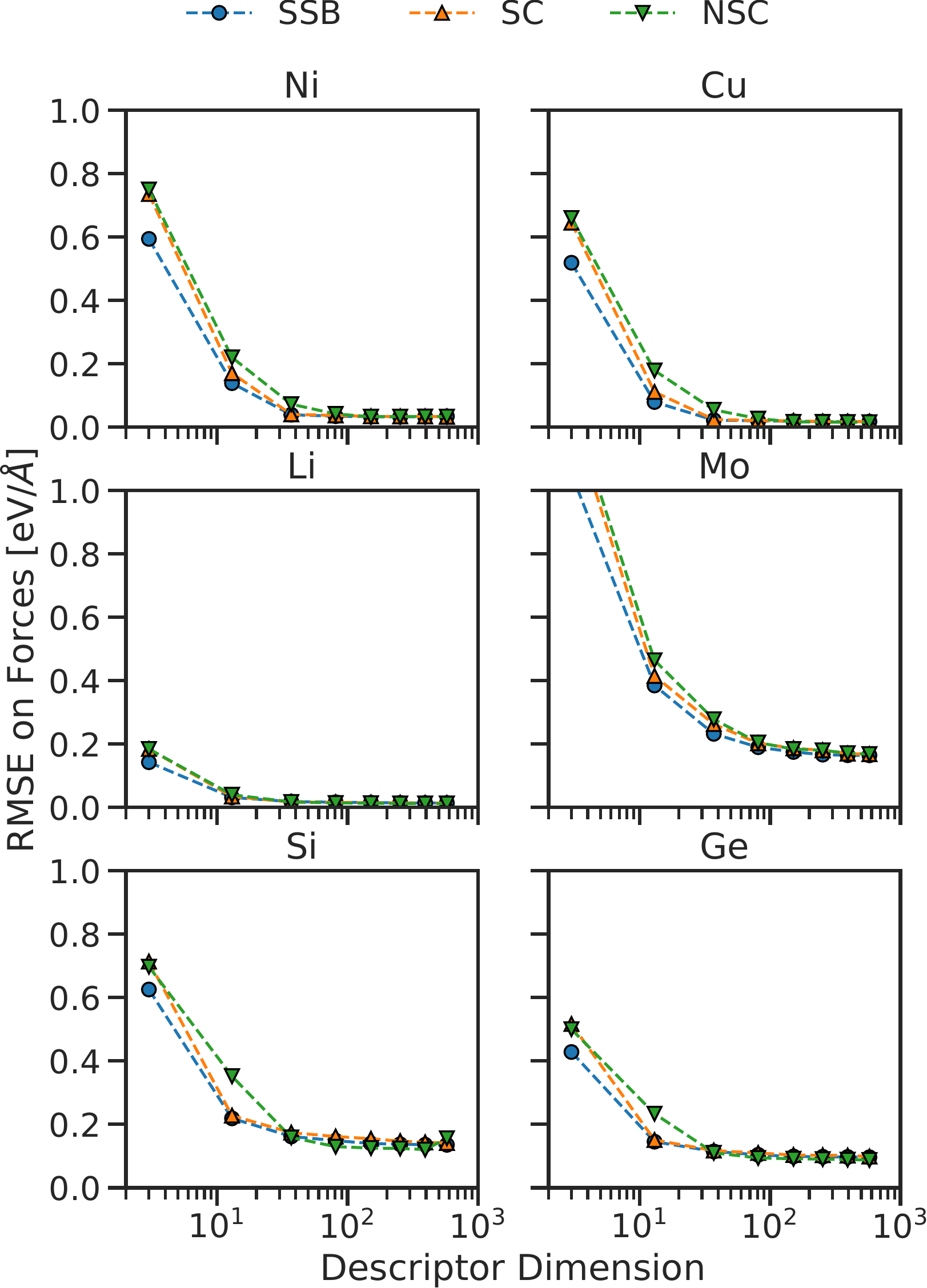}
\caption{RMSE on forces incurred by our RR potential trained and tested on data from Ref.~\onlinecite{zuo2020performance} as a function of the number of features in the representation, using $n_{MAX}$ = $2, \dots, 8$, and $l_{MAX}$ = $n_{MAX}+2$.
Color coding refers to the radial basis functions as in Fig.~\ref{fig:QM9}.}
\label{fig:zuo2020_id_vs_rmse_forces}
\end{figure}
\begin{figure}
\includegraphics[width=8.5cm]{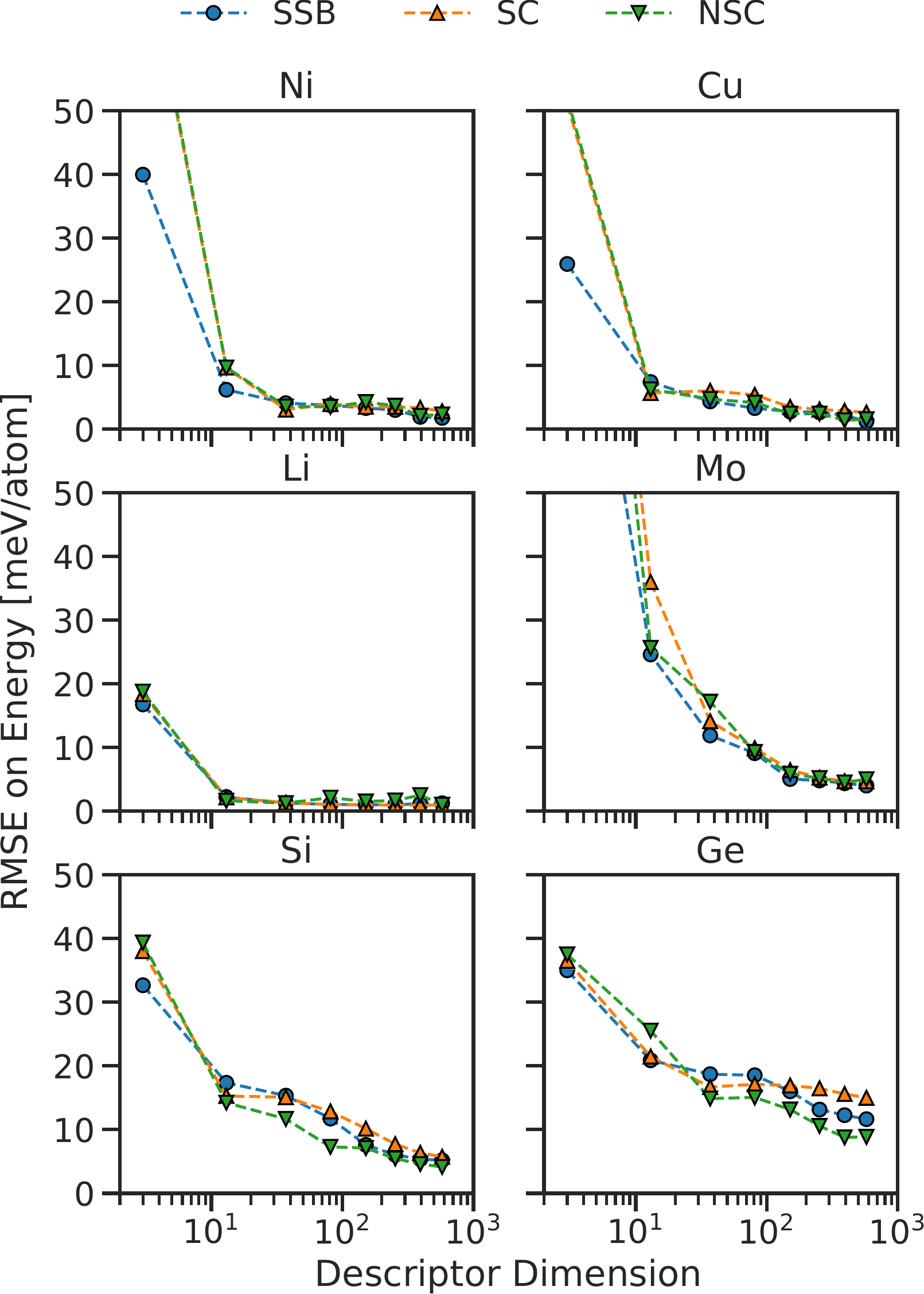}
\caption{RMSE on atomic energies incurred by our RR potential trained and tested on data from Ref.~\onlinecite{zuo2020performance} as a function of the number of features in the representation, using $n_{MAX}$ = $2, \dots, 8$, and $l_{MAX}$ = $n_{MAX} + 2$.
Color coding refers to the radial basis functions as in Fig.~\ref{fig:QM9}.}
\label{fig:zuo2020_id_vs_rmse_energies}
\end{figure}
Similarly to the case of the QM9 dataset, we look at the interplay between number of radial and angular basis employed, and corresponding fitting accuracy.
Figs.~\ref{fig:zuo2020_id_vs_rmse_forces} and ~\ref{fig:zuo2020_id_vs_rmse_energies} report the root mean squared error (RMSE) incurred by the proposed ML framework on forces and on energies, respectively, in each system, as a function of the descriptor's dimension.
An increase in the descriptor extrinsic dimension corresponds to a decrease in the model RMSE, as expected.
For most systems, the RMSE does reach a plateau around a descriptor's dimension of $10^2$, indicating that more compact, and thus more computationally efficient, basis sets can be employed with negligible accuracy loss.
These trends are in agreement with ones previously reported in the literature for other formulations of the local atomic density representations.\cite{Jinnouchi2020} 
In particular, descriptors employing $n_{MAX} = 5$, $l_{MAX} = 7$ incur in RMSEs on forces and energies that are, on average, respectively 0.003 $\pm$ 0.003 eV/$\text{\AA}$ and 1.77 $\pm$ 1.41 meV/atom higher than the ones incurred by the larger descriptor.
In turn, using $n_{MAX} = 5$, $l_{MAX} = 7$ is approximately 4 times faster than $n_{MAX} = 8$, $l_{MAX}=10$.
Differently to the case of the QM9 dataset, in Figs.~\ref{fig:zuo2020_id_vs_rmse_forces} and ~\ref{fig:zuo2020_id_vs_rmse_energies}, the difference in performance of the descriptors employing the three radial basis sets is marginal, even for small descriptor sizes.\\

The current Python implementation of the algorithm favours code interpretability over efficiency.
For a thorough discussion on the computational speed of the ACE framework, we refer the interested reader to Ref.~ \onlinecite{lysogorskiy2021performant}, where it was shown that an efficient C++ implementation of this representation generally leads to predictions whose accuracy and speed are both highly competitive with other state-of-the-art methods.

\section{\label{sec:Dimension}Descriptor Compression}
We have showcased how the use of efficient local atomic environment descriptors can yield a satisfying prediction accuracy, even when employing linear, and thus computationally cheap, regression algorithms.
Nonetheless, the descriptors employed up to this point contain hundreds to thousands of elements, and the question of whether such a large number of variables is really necessary to describe the data naturally arises.
Indeed widely-employed local atomic environment descriptors, such as the smooth overlap of atomic positions (SOAP) and atomic symmetry functions (ASF), can be compressed without loss of accuracy for the case of Gaussian process FFs and artificial neural networks FFs, respectively.~\cite{glielmo2021ranking}
Here, we address this question by applying two different techniques, principal component analysis (PCA) and least absolute shrinkage and selection operator (LASSO) regression, to reduce the dimension of the descriptors employed for energy-only fitting on the 6 single-element datasets analysed in Sec.~\ref{subsec:Zuo}.
For all the six datasets, we compute descriptors $\mathbf{Q}$ using $n_{MAX}$ = 8 and $l_{MAX}$ = 8, we employ the SSB radial basis function, and we avoid augmenting the descriptor with the element-wise square of each element to simplify the analysis of the results.
Figures mirroring the ones shown in the next sections, but for the case of SC and NSC radial basis functions, can be found in the SM, Sections 3 and 4, respectively.
\\

\subsection{PCA Dimensionality Reduction}
\begin{figure}[t!]
    \includegraphics[width=8cm]{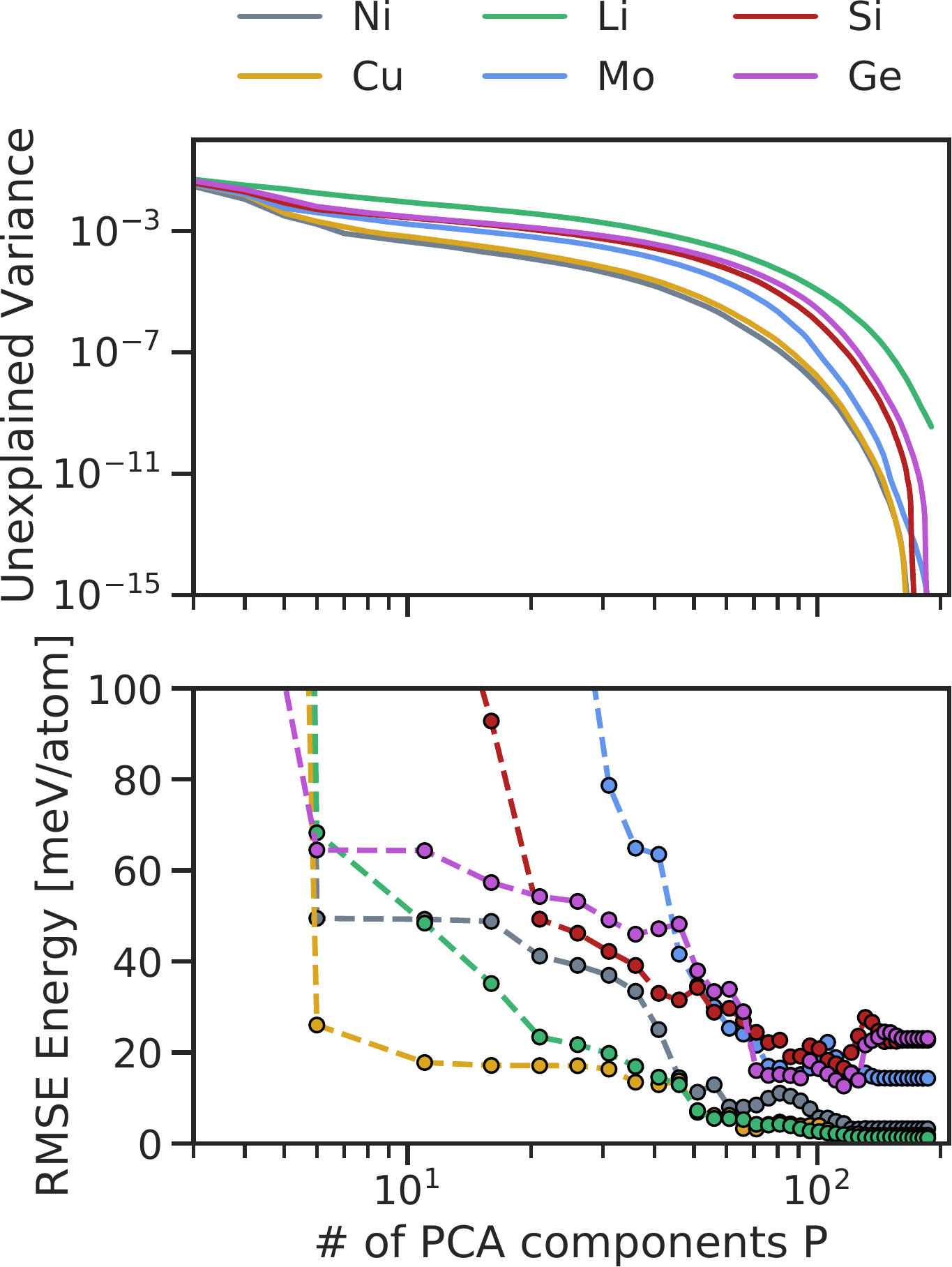}
     \caption{Top panel: data unexplained variance as a function of the number of PCA components accounted for.
     Bottom panel: RMSE on energies incurred by RR potentials employing the reduced descriptor $\mathbf{Q}^{PCA}_{P}$ on the validation set, as a function of the number of PCA components $P$.}
    \label{fig:pca}
\end{figure}

PCA is a well-known data analysis algorithm \cite{Pearson1901}
, often used to draw low-dimensional projections of high-dimensional objects, such as the features deriving from local density representations. \cite{cubuk2017representations}
In a nutshell, PCA fits an ellipsoid to the data (in our case, the descriptors $\mathbf{Q}$), therefore allowing for the identification of the directions of highest variance in the dataset.
Dimensionality reduction can then be performed by employing only the projections of the original data $\mathbf{Q}$ on the $P$ orthogonal directions displaying the highest variance of the aforementioned ellipsoid.
The reduced descriptor is therefore obtained as:
\begin{equation}
\mathbf{Q}^{PCA}_{P} = \mathbf{Q}\cdot\mathbf{C}_P,
\label{eq:PCA_descriptor}
\end{equation}
where $\mathbf{C}_P$ is a matrix with the $P$ directions of maximum variance of the data as columns.
In the top panel of Fig.\ref{fig:pca}, we showcase the fraction of variance of the descriptors $\mathbf{Q}$  which is not explained by the reduced descriptor $\mathbf{Q}^{PCA}_{P}$ as a function of the number $P$ of PCA components used, in a log-log scale.
We notice that around $P\sim80$, for all systems, the curve of unexplained variance sharply changes slope, indicating that the inclusion of components over $P\sim80$ in the reduced vector $\mathbf{Q}^{PCA}_{P}$ will yield a negligible improvement of explained variance.
This is confirmed by the bottom panel of Fig.\ref{fig:pca}, where we report the RMSE on the validation energy prediction of the RR potential employing reduced descriptors $\mathbf{Q}^{PCA}_{P}$ containing $P$ PCA components (solid lines and circles), and of the non-reduced descriptor $\mathbf{Q}$ (crosses), as a function of $P$.
For all elements, the RMSEs have a minimum around $P\sim80$.
Moreover, for the elements displaying a higher RMSE, namely Mo, Si, and Ge, the RMSE increases for $P>80$; this suggests that the inclusion of components beyond $P\sim80$ introduces noise in the descriptor, lowering the validation accuracy. \\
\begin{figure}[t!]
    \includegraphics[width=7cm]{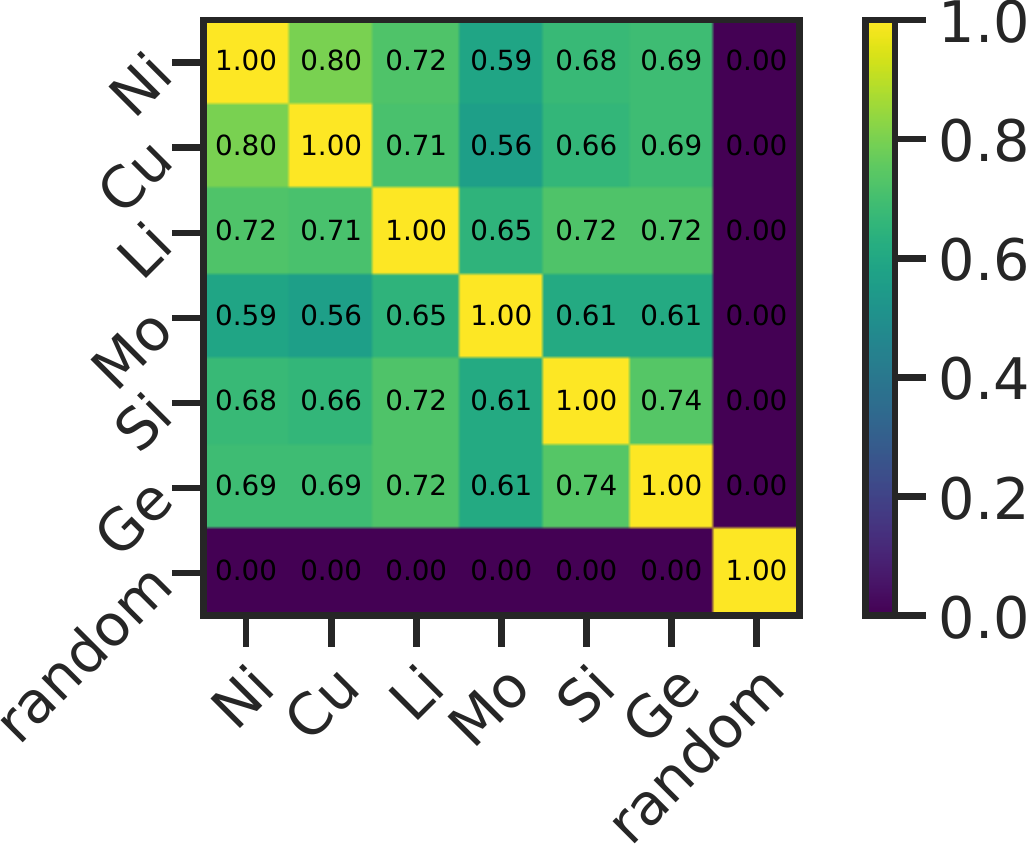}
    \caption{Heatmap displaying the fraction of dimensions shared by the sub-spaces generated by the first 80 PCA-selected directions of the descriptors $\mathbf{Q}$ among couples of single-element datasets.
    The random label indicates a sub-space generated by taking 80 random orthogonal vectors in the space of the $\mathbf{Q}$ vectors.}
    \label{fig:pca_shared_dimensions}
\end{figure}

Using PCA feature selection, we are therefore able to construct a 80-dimensional descriptor that, for each material, %when tested for energy prediction on a validation set,
performs on-par with, and sometimes better than, the full 360-dimensional descriptor.
To investigate the similarity between the matrices $\mathbf{C}_P$ among the six datasets, we look at the dimension of the intersection between the sub-spaces defined by the rows of $\mathbf{C}_P$; details on this procedure are available in the SM, Section 5.
In Fig.~\ref{fig:pca_shared_dimensions}, we report the fraction of shared dimensions between the sub-spaces defined by $\mathbf{C}_P$ with $P=80$.
Elements on the diagonal are 1, as a sub-space shares all of its dimensions with itself, while elements on the last row and column are 0, as these report the fraction of shared dimensions between the sub-spaces defined by $\mathbf{C}_P$ (with $P=80$) and a sub-space defined by 80 360-dimensional randomly generated orthogonal vectors.
The off-diagonal elements of all but the last rows have a mean value of 0.69, indicating that, on average, 55 of the 80 dimensions of the matrix $\mathbf{C}_P$ are shared among couples of single-element descriptor sets.
Similar results are observed when using SC and NSC basis functions, where the average fraction of dimensions of the matrix $\mathbf{C}_P$ that are shared among couples of single-element descriptor sets is 0.93 and 0.78, respectively, as shown in Sections 3 and 4 of the SM.
This indicates a strong redundancy of the feature selection performed by PCA on the different materials, hinting at an underlying material-agnostic structure in the manifold the $\mathbf{Q}$ vectors live in.

\subsection{LASSO-LARS Feature Selection}
\begin{figure}[t!]
    \includegraphics[width=8cm]{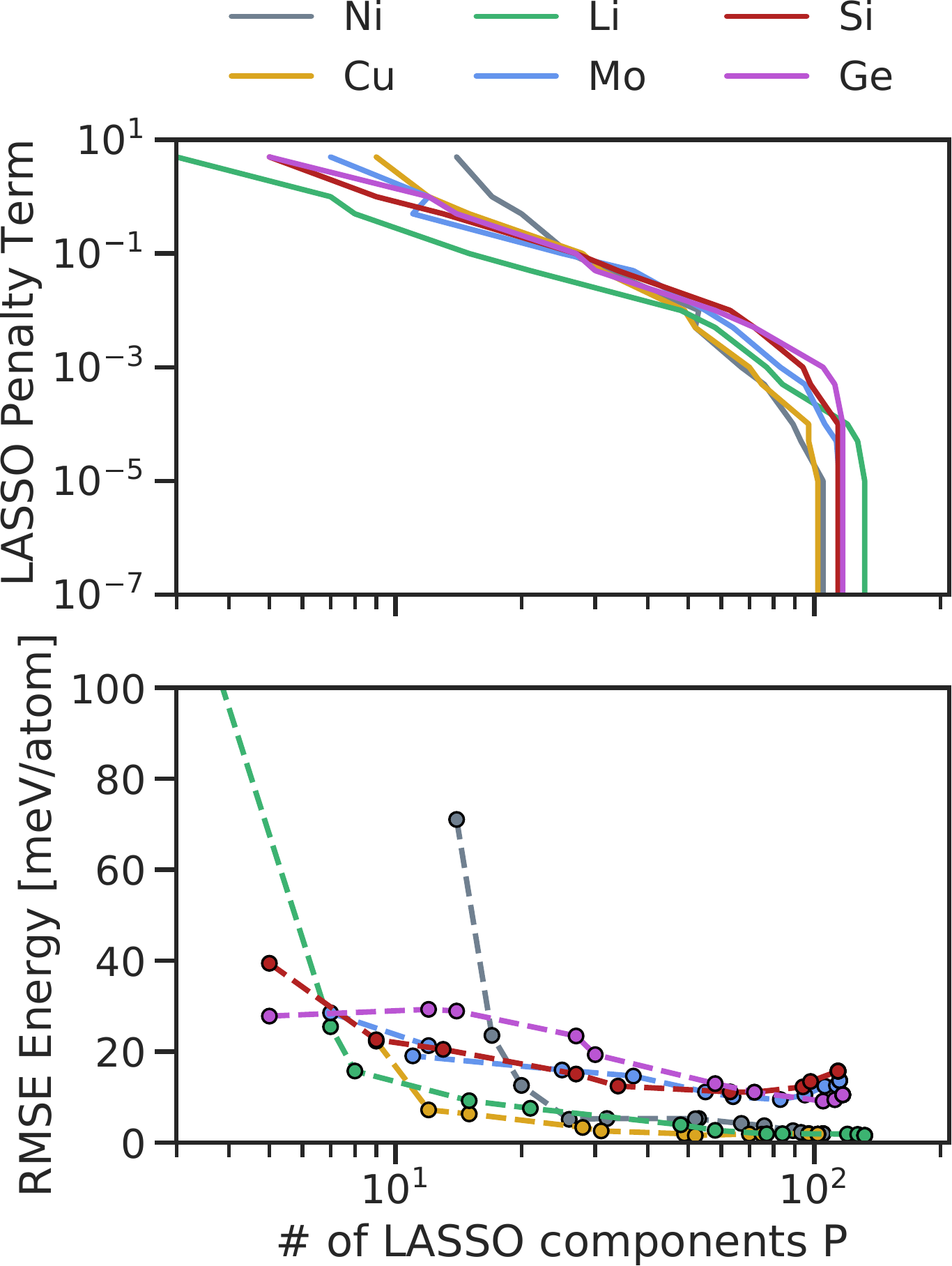}
     \caption{Top panel: LASSO penalty term as a function of the number of LASSO components accounted for.
     Bottom panel: RMSE on energies incurred by RR potentials employing the reduced descriptor $\mathbf{Q}^{LASSO}_{P}$ on the validation set, as a function of the number of LASSO components $P$.}
    \label{fig:lasso}
\end{figure}

We now look at LASSO least angle regression (LARS) as a way to inform the selection of sparse features in the descriptor vector. \cite{efron2004least, Benoit2021}
LASSO LARS is a linear regression algorithm that employs L1 regularization, together with a tunable penalty term.
Here we refer to the features of the original descriptor which have an associated non-zero weight vector in the LASSO LARS linear model as the features that were ``selected'' by the model. \cite{ghiringhelli2015big, ouyang2018sisso, nelson2013compressive}
In the top panel of Fig.~\ref{fig:lasso}, we show the inverse correlation existing between the LASSO LARS penalty term and the number of selected features.
A sharp transition is found for penalty terms around 10$^{-5}$, and after this transition at most 132 out of the 360 features are selected, independently of the material and of the value of the penalty term.
This behaviour strongly suggests that a substantial fraction of the original descriptor contains redundant information, or noise.
This intuition is supported by the lower panel of Fig.~\ref{fig:lasso}, where the validation RMSE on atomic energy displays a minimum when $\sim$~80 features of the original descriptor are employed, for all six materials.

To check whether the same set of features is used for L1 regression across the six materials, we calculate the number of times each descriptor feature is selected by the algorithm when the LASSO penalty term is set to $5\cdot10^{-4}$, i.e. when the average validation RMSE on atomic energy is lowest.
\begin{figure}[t!]
    \includegraphics[width=7cm]{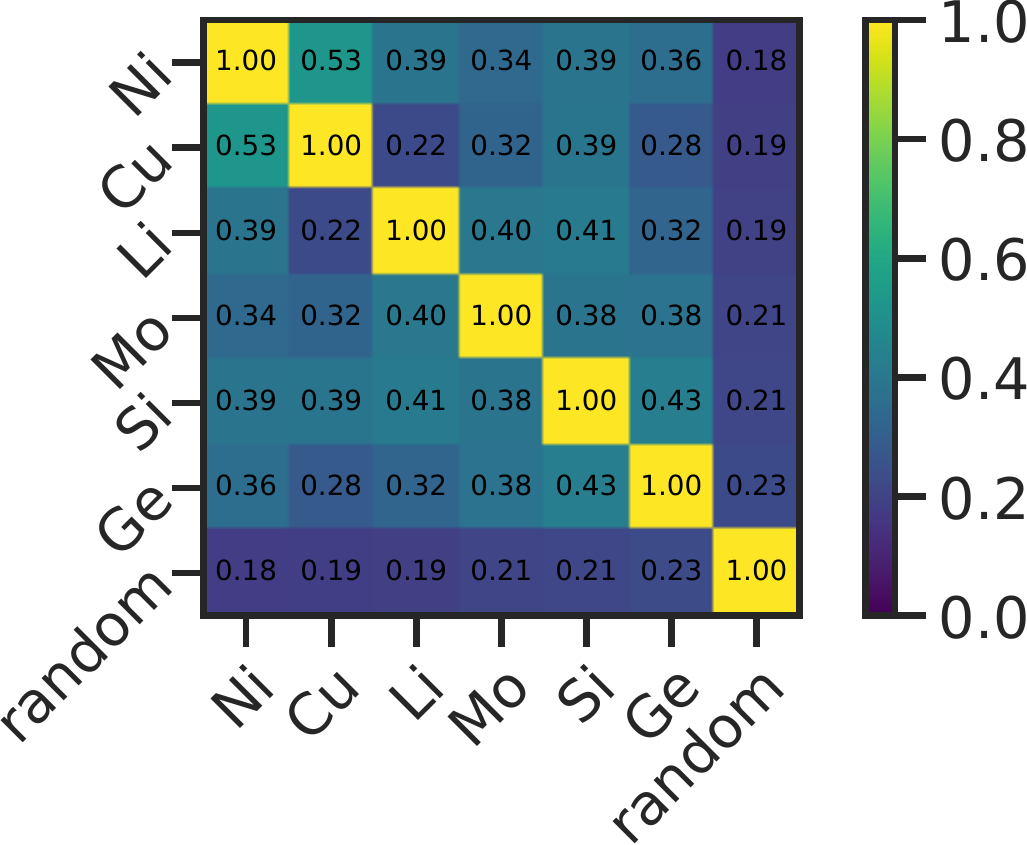}
    \caption{Heatmap displaying the fraction of components selected via LASSO regression that are shared between two materials, with the penalty term set to $5\cdot10^{-4}$.
    The random label indicates a set of 65 randomly selected components out of the available 360.}
    \label{fig:lasso_shared_components}
\end{figure}
In Fig.~\ref{fig:lasso_shared_components} we report the fraction of features selected by the LASSO algorithm shared between each pair of single-element datasets.
On average, 37\% of the LASSO-selected features are shared between each single-element dataset, while only 20\% of features are shared between each single-element set and a randomly-sampled set of 65 features.
Similar results emerge also for the case of SC and NSC radial basis functions, where the percentage of LASSO-selected features that are shared between each single-element dataset is 42\% and 32\%, respectively (SM, Sections C and D, respectively).
The above observations align with the outcome for PCA, and indicate that the number of informative features is much smaller than the dimensionality of the descriptor, and that these features are, at least partially, shared between different materials.
These results, together with the observations drawn from the PCA dimensionality reduction analysis, suggest the possibility to construct an efficient, low-dimensional descriptor, that captures the most relevant features of a local atomic environment.

\section{\label{sec:benchmarks}Conclusions}
We systematically probe the accuracy of a 3-body representation deriving from the ``atomic cluster expansion'' to predict atomic forces and formation energies in solids and molecules.
We furthermore expand the ``atomic cluster representation'' descriptor by employing Bessel polynomials as radial basis function sets, and show that they often display better accuracy than non-scaled Chebyshev and scaled Chebyshev polynomials, when a low number of radial basis functions is employed.
We demonstrate that this representation, coupled with a simple linear regression algorithm, yields a satisfactory prediction accuracy on the QM9 dataset \cite{Rupp2012, Blum2009}, and an accuracy on-par with other state-of-the-art representations and statistical learning methods for six single-element datasets \cite{zuo2020performance}.\\

In second instance we focus on methods to reduce the dimension of the representation.
We consider both a dimensionality reduction scheme (PCA), and a regression algorithm encompassing a feature selection mechanism (LASSO LARS).
We study the interplay between accuracy and representation dimensionality in the database of Ref.~\onlinecite{zuo2020performance}, which comprises FCC metals, BCC metals, and group IV semiconductors.
We find that it is possible to obtain more compact local atomic environment descriptors with no loss in accuracy. 
Furthermore, we find that there exists an ideal number of PCA components or LASSO LARS selected features for which the accuracy in the prediction actually improves, while yielding a four-fold decrease in the dimension of the descriptor.
We then study the structure of the representation resulting from both PCA dimensionality reduction and LASSO LARS  feature selection. 
We find that more than 64\% of the first 80 directions of maximum variance of the descriptors are shared between each pair of single-element datasets.
Similarly, we observe that several descriptor features  are relevant (according to a LASSO LARS selection) for the representation of solids of different nature.
While this result was drawn from databases containing only hundreds of structures, these comprised of elements with diverse chemistry.
In turn, we envision that our approach 
%These results 
could inform the design of extremely compact and fast to compute, yet informative, atomic environment descriptors, in a material-agnostic fashion.

\section*{Supplementary material}
A discussion on how well the Bessel, scaled Chebyshev, and Chebyshev radial basis sets approximate a Dirac delta function,  details on the accuracy trade-off between radial and angular components of the employed local atomic density descriptor, and on the computation of the dimension of the intersection of two sub-spaces, and can be found in the SM.
Furthermore, the SM contains the same plots shown in Figures~\ref{fig:pca}, \ref{fig:pca_shared_dimensions}, \ref{fig:lasso}, and \ref{fig:lasso_shared_components}, but for the case of SC and NSC radial basis functions.
\section*{Acknowledgements}
CZ, AG, and SdG gratefully acknowledge support by the European Union’s Horizon 2020 research and innovation program (Grant No. 824143, MaX `MAterials design at the eXascale' Centre of Excellence).
\section*{Data availability}
The package for training ridge regression potentials is freely available under Apache 2.0 license at: https://github.com/ClaudioZeni/Raffy.
The QM9 dataset is freely available at https://doi.org/10.6084/m9.figshare.c.978904.v5~ \cite{Ramakrishnan_2014}.
The single-element materials dataset is freely available in the data directory at  https://github.com/materialsvirtuallab/mlearn~ \cite{zuo2020performance}.

\section*{Bibliography}

\bibliography{main_arxiv}% Produces the bibliography via BibTeX.

\begin{thebibliography}{55}%
\makeatletter
\providecommand \@ifxundefined [1]{%
 \@ifx{#1\undefined}
}%
\providecommand \@ifnum [1]{%
 \ifnum #1\expandafter \@firstoftwo
 \else \expandafter \@secondoftwo
 \fi
}%
\providecommand \@ifx [1]{%
 \ifx #1\expandafter \@firstoftwo
 \else \expandafter \@secondoftwo
 \fi
}%
\providecommand \natexlab [1]{#1}%
\providecommand \enquote  [1]{``#1''}%
\providecommand \bibnamefont  [1]{#1}%
\providecommand \bibfnamefont [1]{#1}%
\providecommand \citenamefont [1]{#1}%
\providecommand \href@noop [0]{\@secondoftwo}%
\providecommand \href [0]{\begingroup \@sanitize@url \@href}%
\providecommand \@href[1]{\@@startlink{#1}\@@href}%
\providecommand \@@href[1]{\endgroup#1\@@endlink}%
\providecommand \@sanitize@url [0]{\catcode `\\12\catcode `\$12\catcode
  `\&12\catcode `\#12\catcode `\^12\catcode `\_12\catcode `\%12\relax}%
\providecommand \@@startlink[1]{}%
\providecommand \@@endlink[0]{}%
\providecommand \url  [0]{\begingroup\@sanitize@url \@url }%
\providecommand \@url [1]{\endgroup\@href {#1}{\urlprefix }}%
\providecommand \urlprefix  [0]{URL }%
\providecommand \Eprint [0]{\href }%
\providecommand \doibase [0]{http://dx.doi.org/}%
\providecommand \selectlanguage [0]{\@gobble}%
\providecommand \bibinfo  [0]{\@secondoftwo}%
\providecommand \bibfield  [0]{\@secondoftwo}%
\providecommand \translation [1]{[#1]}%
\providecommand \BibitemOpen [0]{}%
\providecommand \bibitemStop [0]{}%
\providecommand \bibitemNoStop [0]{.\EOS\space}%
\providecommand \EOS [0]{\spacefactor3000\relax}%
\providecommand \BibitemShut  [1]{\csname bibitem#1\endcsname}%
\let\auto@bib@innerbib\@empty
%</preamble>
\bibitem [{\citenamefont {Smith}\ \emph {et~al.}(2017)\citenamefont {Smith},
  \citenamefont {Isayev},\ and\ \citenamefont {Roitberg}}]{Smith2017}%
  \BibitemOpen
  \bibfield  {author} {\bibinfo {author} {\bibfnamefont {J.~S.}\ \bibnamefont
  {Smith}}, \bibinfo {author} {\bibfnamefont {O.}~\bibnamefont {Isayev}}, \
  and\ \bibinfo {author} {\bibfnamefont {A.~E.}\ \bibnamefont {Roitberg}},\
  }\href {\doibase 10.1039/C6SC05720A} {\bibfield  {journal} {\bibinfo
  {journal} {Chemical Science}\ }\textbf {\bibinfo {volume} {8}} (\bibinfo
  {year} {2017}),\ 10.1039/C6SC05720A}\BibitemShut {NoStop}%
\bibitem [{\citenamefont {Artrith}(2019)}]{Artrith2019}%
  \BibitemOpen
  \bibfield  {author} {\bibinfo {author} {\bibfnamefont {N.}~\bibnamefont
  {Artrith}},\ }\href {\doibase 10.1088/2515-7655/ab2060} {\bibfield  {journal}
  {\bibinfo  {journal} {Journal of Physics: Energy}\ }\textbf {\bibinfo
  {volume} {1}} (\bibinfo {year} {2019}),\
  10.1088/2515-7655/ab2060}\BibitemShut {NoStop}%
\bibitem [{\citenamefont {Schleder}\ \emph {et~al.}(2019)\citenamefont
  {Schleder}, \citenamefont {Padilha}, \citenamefont {Acosta}, \citenamefont
  {Costa},\ and\ \citenamefont {Fazzio}}]{Schleder2019}%
  \BibitemOpen
  \bibfield  {author} {\bibinfo {author} {\bibfnamefont {G.~R.}\ \bibnamefont
  {Schleder}}, \bibinfo {author} {\bibfnamefont {A.~C.~M.}\ \bibnamefont
  {Padilha}}, \bibinfo {author} {\bibfnamefont {C.~M.}\ \bibnamefont {Acosta}},
  \bibinfo {author} {\bibfnamefont {M.}~\bibnamefont {Costa}}, \ and\ \bibinfo
  {author} {\bibfnamefont {A.}~\bibnamefont {Fazzio}},\ }\href {\doibase
  10.1088/2515-7639/ab084b} {\bibfield  {journal} {\bibinfo  {journal} {Journal
  of Physics: Materials}\ }\textbf {\bibinfo {volume} {2}} (\bibinfo {year}
  {2019}),\ 10.1088/2515-7639/ab084b}\BibitemShut {NoStop}%
\bibitem [{\citenamefont {Paruzzo}\ \emph {et~al.}(2018)\citenamefont
  {Paruzzo}, \citenamefont {Hofstetter}, \citenamefont {Musil}, \citenamefont
  {De}, \citenamefont {Ceriotti},\ and\ \citenamefont {Emsley}}]{Paruzzo2018}%
  \BibitemOpen
  \bibfield  {author} {\bibinfo {author} {\bibfnamefont {F.~M.}\ \bibnamefont
  {Paruzzo}}, \bibinfo {author} {\bibfnamefont {A.}~\bibnamefont {Hofstetter}},
  \bibinfo {author} {\bibfnamefont {F.}~\bibnamefont {Musil}}, \bibinfo
  {author} {\bibfnamefont {S.}~\bibnamefont {De}}, \bibinfo {author}
  {\bibfnamefont {M.}~\bibnamefont {Ceriotti}}, \ and\ \bibinfo {author}
  {\bibfnamefont {L.}~\bibnamefont {Emsley}},\ }\href {\doibase
  10.1038/s41467-018-06972-x} {\bibfield  {journal} {\bibinfo  {journal}
  {Nature Communications}\ }\textbf {\bibinfo {volume} {9}} (\bibinfo {year}
  {2018}),\ 10.1038/s41467-018-06972-x}\BibitemShut {NoStop}%
\bibitem [{\citenamefont {Gupta}\ \emph {et~al.}(2021)\citenamefont {Gupta},
  \citenamefont {Chakraborty},\ and\ \citenamefont {Ramakrishnan}}]{Gupta2021}%
  \BibitemOpen
  \bibfield  {author} {\bibinfo {author} {\bibfnamefont {A.}~\bibnamefont
  {Gupta}}, \bibinfo {author} {\bibfnamefont {S.}~\bibnamefont {Chakraborty}},
  \ and\ \bibinfo {author} {\bibfnamefont {R.}~\bibnamefont {Ramakrishnan}},\
  }\href {\doibase 10.1088/2632-2153/abe347} {\bibfield  {journal} {\bibinfo
  {journal} {Machine Learning: Science and Technology}\ } (\bibinfo {year}
  {2021}),\ 10.1088/2632-2153/abe347}\BibitemShut {NoStop}%
\bibitem [{\citenamefont {Wilkins}\ \emph {et~al.}(2019)\citenamefont
  {Wilkins}, \citenamefont {Grisafi}, \citenamefont {Yang}, \citenamefont
  {Lao}, \citenamefont {DiStasio},\ and\ \citenamefont
  {Ceriotti}}]{wilkins2019}%
  \BibitemOpen
  \bibfield  {author} {\bibinfo {author} {\bibfnamefont {D.~M.}\ \bibnamefont
  {Wilkins}}, \bibinfo {author} {\bibfnamefont {A.}~\bibnamefont {Grisafi}},
  \bibinfo {author} {\bibfnamefont {Y.}~\bibnamefont {Yang}}, \bibinfo {author}
  {\bibfnamefont {K.~U.}\ \bibnamefont {Lao}}, \bibinfo {author} {\bibfnamefont
  {R.~A.}\ \bibnamefont {DiStasio}}, \ and\ \bibinfo {author} {\bibfnamefont
  {M.}~\bibnamefont {Ceriotti}},\ }\href {\doibase 10.1073/pnas.1816132116}
  {\bibfield  {journal} {\bibinfo  {journal} {Proceedings of the National
  Academy of Sciences of the United States of America}\ }\textbf {\bibinfo
  {volume} {116}} (\bibinfo {year} {2019}),\
  10.1073/pnas.1816132116}\BibitemShut {NoStop}%
\bibitem [{\citenamefont {Rauer}\ and\ \citenamefont
  {Bereau}(2020)}]{Rauer2020}%
  \BibitemOpen
  \bibfield  {author} {\bibinfo {author} {\bibfnamefont {C.}~\bibnamefont
  {Rauer}}\ and\ \bibinfo {author} {\bibfnamefont {T.}~\bibnamefont {Bereau}},\
  }\href {\doibase 10.1063/5.0012230} {\bibfield  {journal} {\bibinfo
  {journal} {Journal of Chemical Physics}\ }\textbf {\bibinfo {volume} {153}},\
  \bibinfo {pages} {014101} (\bibinfo {year} {2020})},\ \Eprint
  {http://arxiv.org/abs/2007.00407} {arXiv:2007.00407} \BibitemShut {NoStop}%
\bibitem [{\citenamefont {Wu}\ \emph {et~al.}(2018)\citenamefont {Wu},
  \citenamefont {Ramsundar}, \citenamefont {Feinberg}, \citenamefont {Gomes},
  \citenamefont {Geniesse}, \citenamefont {Pappu}, \citenamefont {Leswing},\
  and\ \citenamefont {Pande}}]{Wu2018}%
  \BibitemOpen
  \bibfield  {author} {\bibinfo {author} {\bibfnamefont {Z.}~\bibnamefont
  {Wu}}, \bibinfo {author} {\bibfnamefont {B.}~\bibnamefont {Ramsundar}},
  \bibinfo {author} {\bibfnamefont {E.~N.}\ \bibnamefont {Feinberg}}, \bibinfo
  {author} {\bibfnamefont {J.}~\bibnamefont {Gomes}}, \bibinfo {author}
  {\bibfnamefont {C.}~\bibnamefont {Geniesse}}, \bibinfo {author}
  {\bibfnamefont {A.~S.}\ \bibnamefont {Pappu}}, \bibinfo {author}
  {\bibfnamefont {K.}~\bibnamefont {Leswing}}, \ and\ \bibinfo {author}
  {\bibfnamefont {V.}~\bibnamefont {Pande}},\ }\href {\doibase
  10.1039/c7sc02664a} {\bibfield  {journal} {\bibinfo  {journal} {Chemical
  Science}\ }\textbf {\bibinfo {volume} {9}},\ \bibinfo {pages} {513} (\bibinfo
  {year} {2018})},\ \Eprint {http://arxiv.org/abs/1703.00564}
  {arXiv:1703.00564} \BibitemShut {NoStop}%
\bibitem [{\citenamefont {Axelrod}\ and\ \citenamefont
  {Gomez-Bombarelli}(2020)}]{Axelrod2020}%
  \BibitemOpen
  \bibfield  {author} {\bibinfo {author} {\bibfnamefont {S.}~\bibnamefont
  {Axelrod}}\ and\ \bibinfo {author} {\bibfnamefont {R.}~\bibnamefont
  {Gomez-Bombarelli}},\ }\href {http://arxiv.org/abs/2006.05531} {\bibfield
  {journal} {\bibinfo  {journal} {arXiv}\ } (\bibinfo {year} {2020})},\ \Eprint
  {http://arxiv.org/abs/2006.05531} {arXiv:2006.05531} \BibitemShut {NoStop}%
\bibitem [{\citenamefont {J{\"{a}}ger}\ \emph {et~al.}(2018)\citenamefont
  {J{\"{a}}ger}, \citenamefont {Morooka}, \citenamefont {Canova}, \citenamefont
  {Himanen},\ and\ \citenamefont {Foster}}]{Jager2018}%
  \BibitemOpen
  \bibfield  {author} {\bibinfo {author} {\bibfnamefont {M.~O.~J.}\
  \bibnamefont {J{\"{a}}ger}}, \bibinfo {author} {\bibfnamefont {E.~V.}\
  \bibnamefont {Morooka}}, \bibinfo {author} {\bibfnamefont {F.~F.}\
  \bibnamefont {Canova}}, \bibinfo {author} {\bibfnamefont {L.}~\bibnamefont
  {Himanen}}, \ and\ \bibinfo {author} {\bibfnamefont {A.~S.}\ \bibnamefont
  {Foster}},\ }\href {http://dx.doi.org/10.1038/s41524-018-0096-5} {\bibfield
  {journal} {\bibinfo  {journal} {npj Computational Materials}\ } (\bibinfo
  {year} {2018})}\BibitemShut {NoStop}%
\bibitem [{\citenamefont {Meyer}\ \emph {et~al.}(2018)\citenamefont {Meyer},
  \citenamefont {Sawatlon}, \citenamefont {Heinen}, \citenamefont {{Von
  Lilienfeld}},\ and\ \citenamefont {Corminboeuf}}]{Meyer2018}%
  \BibitemOpen
  \bibfield  {author} {\bibinfo {author} {\bibfnamefont {B.}~\bibnamefont
  {Meyer}}, \bibinfo {author} {\bibfnamefont {B.}~\bibnamefont {Sawatlon}},
  \bibinfo {author} {\bibfnamefont {S.}~\bibnamefont {Heinen}}, \bibinfo
  {author} {\bibfnamefont {O.~A.}\ \bibnamefont {{Von Lilienfeld}}}, \ and\
  \bibinfo {author} {\bibfnamefont {C.}~\bibnamefont {Corminboeuf}},\ }\href
  {\doibase 10.1039/c8sc01949e} {\bibfield  {journal} {\bibinfo  {journal}
  {Chemical Science}\ }\textbf {\bibinfo {volume} {9}},\ \bibinfo {pages}
  {7069} (\bibinfo {year} {2018})}\BibitemShut {NoStop}%
\bibitem [{\citenamefont {Gu}\ \emph {et~al.}(2020)\citenamefont {Gu},
  \citenamefont {Noh}, \citenamefont {Kim}, \citenamefont {Back}, \citenamefont
  {Ulissi},\ and\ \citenamefont {Jung}}]{Gu2020}%
  \BibitemOpen
  \bibfield  {author} {\bibinfo {author} {\bibfnamefont {G.~H.}\ \bibnamefont
  {Gu}}, \bibinfo {author} {\bibfnamefont {J.}~\bibnamefont {Noh}}, \bibinfo
  {author} {\bibfnamefont {S.}~\bibnamefont {Kim}}, \bibinfo {author}
  {\bibfnamefont {S.}~\bibnamefont {Back}}, \bibinfo {author} {\bibfnamefont
  {Z.}~\bibnamefont {Ulissi}}, \ and\ \bibinfo {author} {\bibfnamefont
  {Y.}~\bibnamefont {Jung}},\ }\href
  {https://doi.org/10.1021/acs.jpclett.0c00634} {\bibfield  {journal} {\bibinfo
   {journal} {The journal of physical chemistry letters}\ }\textbf {\bibinfo
  {volume} {11}},\ \bibinfo {pages} {3185} (\bibinfo {year}
  {2020})}\BibitemShut {NoStop}%
\bibitem [{\citenamefont {Rossi}\ \emph {et~al.}(2020)\citenamefont {Rossi},
  \citenamefont {Jur{\'{a}}skov{\'{a}}}, \citenamefont {Wischert},
  \citenamefont {Garel}, \citenamefont {Corminb{\oe}uf},\ and\ \citenamefont
  {Ceriotti}}]{Rossi2020}%
  \BibitemOpen
  \bibfield  {author} {\bibinfo {author} {\bibfnamefont {K.}~\bibnamefont
  {Rossi}}, \bibinfo {author} {\bibfnamefont {V.}~\bibnamefont
  {Jur{\'{a}}skov{\'{a}}}}, \bibinfo {author} {\bibfnamefont {R.}~\bibnamefont
  {Wischert}}, \bibinfo {author} {\bibfnamefont {L.}~\bibnamefont {Garel}},
  \bibinfo {author} {\bibfnamefont {C.}~\bibnamefont {Corminb{\oe}uf}}, \ and\
  \bibinfo {author} {\bibfnamefont {M.}~\bibnamefont {Ceriotti}},\ }\href
  {\doibase 10.1021/acs.jctc.0c00362} {\bibfield  {journal} {\bibinfo
  {journal} {Journal of Chemical Theory and Computation}\ }\textbf {\bibinfo
  {volume} {16}} (\bibinfo {year} {2020}),\
  10.1021/acs.jctc.0c00362}\BibitemShut {NoStop}%
\bibitem [{\citenamefont {Yang}\ \emph {et~al.}(2020)\citenamefont {Yang},
  \citenamefont {Bonati}, \citenamefont {Polino},\ and\ \citenamefont
  {Parrinello}}]{Yang2020}%
  \BibitemOpen
  \bibfield  {author} {\bibinfo {author} {\bibfnamefont {M.}~\bibnamefont
  {Yang}}, \bibinfo {author} {\bibfnamefont {L.}~\bibnamefont {Bonati}},
  \bibinfo {author} {\bibfnamefont {D.}~\bibnamefont {Polino}}, \ and\ \bibinfo
  {author} {\bibfnamefont {M.}~\bibnamefont {Parrinello}},\ }\href
  {http://arxiv.org/abs/2011.11455} {\bibfield  {journal} {\bibinfo  {journal}
  {arXiv}\ } (\bibinfo {year} {2020})},\ \Eprint
  {http://arxiv.org/abs/2011.11455} {arXiv:2011.11455} \BibitemShut {NoStop}%
\bibitem [{\citenamefont {Eckhoff}\ \emph {et~al.}(2020)\citenamefont
  {Eckhoff}, \citenamefont {Sch{\"{o}}newald}, \citenamefont {Risch},
  \citenamefont {Volkert}, \citenamefont {Bl{\"{o}}chl},\ and\ \citenamefont
  {Behler}}]{Eckhoff2020}%
  \BibitemOpen
  \bibfield  {author} {\bibinfo {author} {\bibfnamefont {M.}~\bibnamefont
  {Eckhoff}}, \bibinfo {author} {\bibfnamefont {F.}~\bibnamefont
  {Sch{\"{o}}newald}}, \bibinfo {author} {\bibfnamefont {M.}~\bibnamefont
  {Risch}}, \bibinfo {author} {\bibfnamefont {C.~A.}\ \bibnamefont {Volkert}},
  \bibinfo {author} {\bibfnamefont {P.~E.}\ \bibnamefont {Bl{\"{o}}chl}}, \
  and\ \bibinfo {author} {\bibfnamefont {J.}~\bibnamefont {Behler}},\ }\href
  {\doibase 10.1103/PhysRevB.102.174102} {\bibfield  {journal} {\bibinfo
  {journal} {Physical Review B}\ }\textbf {\bibinfo {volume} {102}},\ \bibinfo
  {pages} {174102} (\bibinfo {year} {2020})},\ \Eprint
  {http://arxiv.org/abs/2007.00327} {arXiv:2007.00327} \BibitemShut {NoStop}%
\bibitem [{\citenamefont {Vandermause}\ \emph {et~al.}(2020)\citenamefont
  {Vandermause}, \citenamefont {Torrisi}, \citenamefont {Batzner},
  \citenamefont {Xie}, \citenamefont {Sun}, \citenamefont {Kolpak},\ and\
  \citenamefont {Kozinsky}}]{vandermause2020fly}%
  \BibitemOpen
  \bibfield  {author} {\bibinfo {author} {\bibfnamefont {J.}~\bibnamefont
  {Vandermause}}, \bibinfo {author} {\bibfnamefont {S.~B.}\ \bibnamefont
  {Torrisi}}, \bibinfo {author} {\bibfnamefont {S.}~\bibnamefont {Batzner}},
  \bibinfo {author} {\bibfnamefont {Y.}~\bibnamefont {Xie}}, \bibinfo {author}
  {\bibfnamefont {L.}~\bibnamefont {Sun}}, \bibinfo {author} {\bibfnamefont
  {A.~M.}\ \bibnamefont {Kolpak}}, \ and\ \bibinfo {author} {\bibfnamefont
  {B.}~\bibnamefont {Kozinsky}},\ }\href {\doibase
  https://doi.org/10.1038/s41524-020-0283-z} {\bibfield  {journal} {\bibinfo
  {journal} {npj Computational Materials}\ }\textbf {\bibinfo {volume} {6}},\
  \bibinfo {pages} {1} (\bibinfo {year} {2020})}\BibitemShut {NoStop}%
\bibitem [{\citenamefont {Zeni}\ \emph {et~al.}(2019)\citenamefont {Zeni},
  \citenamefont {Rossi}, \citenamefont {Glielmo},\ and\ \citenamefont
  {Baletto}}]{Zeni2019}%
  \BibitemOpen
  \bibfield  {author} {\bibinfo {author} {\bibfnamefont {C.}~\bibnamefont
  {Zeni}}, \bibinfo {author} {\bibfnamefont {K.}~\bibnamefont {Rossi}},
  \bibinfo {author} {\bibfnamefont {A.}~\bibnamefont {Glielmo}}, \ and\
  \bibinfo {author} {\bibfnamefont {F.}~\bibnamefont {Baletto}},\ }\href
  {\doibase 10.1080/23746149.2019.1654919} {\bibfield  {journal} {\bibinfo
  {journal} {Advances in Physics: X}\ }\textbf {\bibinfo {volume} {4}},\
  \bibinfo {pages} {1654919} (\bibinfo {year} {2019})},\ \Eprint
  {http://arxiv.org/abs/https://doi.org/10.1080/23746149.2019.1654919}
  {https://doi.org/10.1080/23746149.2019.1654919} \BibitemShut {NoStop}%
\bibitem [{\citenamefont {Sosso}\ \emph {et~al.}(2012)\citenamefont {Sosso},
  \citenamefont {Miceli}, \citenamefont {Caravati}, \citenamefont {Behler},\
  and\ \citenamefont {Bernasconi}}]{Sosso2012}%
  \BibitemOpen
  \bibfield  {author} {\bibinfo {author} {\bibfnamefont {G.~C.}\ \bibnamefont
  {Sosso}}, \bibinfo {author} {\bibfnamefont {G.}~\bibnamefont {Miceli}},
  \bibinfo {author} {\bibfnamefont {S.}~\bibnamefont {Caravati}}, \bibinfo
  {author} {\bibfnamefont {J.}~\bibnamefont {Behler}}, \ and\ \bibinfo {author}
  {\bibfnamefont {M.}~\bibnamefont {Bernasconi}},\ }\href {\doibase
  10.1103/PhysRevB.85.174103} {\bibfield  {journal} {\bibinfo  {journal}
  {Physical Review B - Condensed Matter and Materials Physics}\ }\textbf
  {\bibinfo {volume} {85}},\ \bibinfo {pages} {174103} (\bibinfo {year}
  {2012})},\ \Eprint {http://arxiv.org/abs/1201.2026} {arXiv:1201.2026}
  \BibitemShut {NoStop}%
\bibitem [{\citenamefont {Zeni}\ \emph {et~al.}(2018)\citenamefont {Zeni},
  \citenamefont {Rossi}, \citenamefont {Glielmo}, \citenamefont {Fekete},
  \citenamefont {Gaston}, \citenamefont {Baletto},\ and\ \citenamefont {{De
  Vita}}}]{Zeni2018}%
  \BibitemOpen
  \bibfield  {author} {\bibinfo {author} {\bibfnamefont {C.}~\bibnamefont
  {Zeni}}, \bibinfo {author} {\bibfnamefont {K.}~\bibnamefont {Rossi}},
  \bibinfo {author} {\bibfnamefont {A.}~\bibnamefont {Glielmo}}, \bibinfo
  {author} {\bibfnamefont {{\'{A}}.}~\bibnamefont {Fekete}}, \bibinfo {author}
  {\bibfnamefont {N.}~\bibnamefont {Gaston}}, \bibinfo {author} {\bibfnamefont
  {F.}~\bibnamefont {Baletto}}, \ and\ \bibinfo {author} {\bibfnamefont
  {A.}~\bibnamefont {{De Vita}}},\ }\href
  {https://aip.scitation.org/doi/10.1063/1.5024558} {\bibfield  {journal}
  {\bibinfo  {journal} {Journal of Chemical Physics}\ }\textbf {\bibinfo
  {volume} {148}} (\bibinfo {year} {2018})},\ \Eprint
  {http://arxiv.org/abs/1802.01417} {1802.01417} \BibitemShut {NoStop}%
\bibitem [{\citenamefont {Deringer}\ \emph {et~al.}(2021)\citenamefont
  {Deringer}, \citenamefont {Bernstein}, \citenamefont {Cs{\'{a}}nyi},
  \citenamefont {{Ben Mahmoud}}, \citenamefont {Ceriotti}, \citenamefont
  {Wilson}, \citenamefont {Drabold},\ and\ \citenamefont
  {Elliott}}]{Deringer2021}%
  \BibitemOpen
  \bibfield  {author} {\bibinfo {author} {\bibfnamefont {V.~L.}\ \bibnamefont
  {Deringer}}, \bibinfo {author} {\bibfnamefont {N.}~\bibnamefont {Bernstein}},
  \bibinfo {author} {\bibfnamefont {G.}~\bibnamefont {Cs{\'{a}}nyi}}, \bibinfo
  {author} {\bibfnamefont {C.}~\bibnamefont {{Ben Mahmoud}}}, \bibinfo {author}
  {\bibfnamefont {M.}~\bibnamefont {Ceriotti}}, \bibinfo {author}
  {\bibfnamefont {M.}~\bibnamefont {Wilson}}, \bibinfo {author} {\bibfnamefont
  {D.~A.}\ \bibnamefont {Drabold}}, \ and\ \bibinfo {author} {\bibfnamefont
  {S.~R.}\ \bibnamefont {Elliott}},\ }\href {\doibase
  10.1038/s41586-020-03072-z} {\bibfield  {journal} {\bibinfo  {journal}
  {Nature}\ }\textbf {\bibinfo {volume} {589}},\ \bibinfo {pages} {59}
  (\bibinfo {year} {2021})}\BibitemShut {NoStop}%
\bibitem [{\citenamefont {Zhang}\ \emph {et~al.}(2020)\citenamefont {Zhang},
  \citenamefont {Cs{\'a}nyi},\ and\ \citenamefont {Alf{\`e}}}]{Zhang2020}%
  \BibitemOpen
  \bibfield  {author} {\bibinfo {author} {\bibfnamefont {Z.}~\bibnamefont
  {Zhang}}, \bibinfo {author} {\bibfnamefont {G.}~\bibnamefont {Cs{\'a}nyi}}, \
  and\ \bibinfo {author} {\bibfnamefont {D.}~\bibnamefont {Alf{\`e}}},\ }\href
  {https://doi.org/10.1016/j.gca.2020.03.028} {\bibfield  {journal} {\bibinfo
  {journal} {Geochimica et Cosmochimica Acta}\ }\textbf {\bibinfo {volume}
  {291}},\ \bibinfo {pages} {5} (\bibinfo {year} {2020})}\BibitemShut {NoStop}%
\bibitem [{\citenamefont {Lahey}\ and\ \citenamefont
  {Rowley}(2020)}]{Lahey2020}%
  \BibitemOpen
  \bibfield  {author} {\bibinfo {author} {\bibfnamefont {S.~L.~J.}\
  \bibnamefont {Lahey}}\ and\ \bibinfo {author} {\bibfnamefont {C.~N.}\
  \bibnamefont {Rowley}},\ }\href {\doibase 10.1039/c9sc06017k} {\bibfield
  {journal} {\bibinfo  {journal} {Chemical Science}\ }\textbf {\bibinfo
  {volume} {11}},\ \bibinfo {pages} {2362} (\bibinfo {year}
  {2020})}\BibitemShut {NoStop}%
\bibitem [{\citenamefont {Wang}\ \emph {et~al.}(2019)\citenamefont {Wang},
  \citenamefont {Olsson}, \citenamefont {Wehmeyer}, \citenamefont
  {P{\'{e}}rez}, \citenamefont {Charron}, \citenamefont {{De Fabritiis}},
  \citenamefont {No{\'{e}}},\ and\ \citenamefont {Clementi}}]{Wang2019}%
  \BibitemOpen
  \bibfield  {author} {\bibinfo {author} {\bibfnamefont {J.}~\bibnamefont
  {Wang}}, \bibinfo {author} {\bibfnamefont {S.}~\bibnamefont {Olsson}},
  \bibinfo {author} {\bibfnamefont {C.}~\bibnamefont {Wehmeyer}}, \bibinfo
  {author} {\bibfnamefont {A.}~\bibnamefont {P{\'{e}}rez}}, \bibinfo {author}
  {\bibfnamefont {N.~E.}\ \bibnamefont {Charron}}, \bibinfo {author}
  {\bibfnamefont {G.}~\bibnamefont {{De Fabritiis}}}, \bibinfo {author}
  {\bibfnamefont {F.}~\bibnamefont {No{\'{e}}}}, \ and\ \bibinfo {author}
  {\bibfnamefont {C.}~\bibnamefont {Clementi}},\ }\href {\doibase
  10.1021/acscentsci.8b00913} {\bibfield  {journal} {\bibinfo  {journal} {ACS
  Central Science}\ }\textbf {\bibinfo {volume} {5}},\ \bibinfo {pages} {755}
  (\bibinfo {year} {2019})},\ \Eprint {http://arxiv.org/abs/1812.01736}
  {arXiv:1812.01736} \BibitemShut {NoStop}%
\bibitem [{\citenamefont {Scherer}\ \emph {et~al.}(2020)\citenamefont
  {Scherer}, \citenamefont {Scheid}, \citenamefont {Andrienko},\ and\
  \citenamefont {Bereau}}]{Scherer2020}%
  \BibitemOpen
  \bibfield  {author} {\bibinfo {author} {\bibfnamefont {C.}~\bibnamefont
  {Scherer}}, \bibinfo {author} {\bibfnamefont {R.}~\bibnamefont {Scheid}},
  \bibinfo {author} {\bibfnamefont {D.}~\bibnamefont {Andrienko}}, \ and\
  \bibinfo {author} {\bibfnamefont {T.}~\bibnamefont {Bereau}},\ }\href
  {\doibase 10.1021/acs.jctc.9b01256} {\bibfield  {journal} {\bibinfo
  {journal} {Journal of Chemical Theory and Computation}\ }\textbf {\bibinfo
  {volume} {16}},\ \bibinfo {pages} {3194} (\bibinfo {year}
  {2020})}\BibitemShut {NoStop}%
\bibitem [{\citenamefont {Behler}\ and\ \citenamefont
  {Parrinello}(2007)}]{Behler:2007fe}%
  \BibitemOpen
  \bibfield  {author} {\bibinfo {author} {\bibfnamefont {J.}~\bibnamefont
  {Behler}}\ and\ \bibinfo {author} {\bibfnamefont {M.}~\bibnamefont
  {Parrinello}},\ }\href
  {https://journals.aps.org/prl/abstract/10.1103/PhysRevLett.98.146401}
  {\bibfield  {journal} {\bibinfo  {journal} {Physical Review Letters}\
  }\textbf {\bibinfo {volume} {98}},\ \bibinfo {pages} {146401} (\bibinfo
  {year} {2007})}\BibitemShut {NoStop}%
\bibitem [{\citenamefont {Bart\'ok}\ \emph {et~al.}(2010)\citenamefont
  {Bart\'ok}, \citenamefont {Payne}, \citenamefont {Kondor},\ and\
  \citenamefont {Cs\'anyi}}]{Bartok2010}%
  \BibitemOpen
  \bibfield  {author} {\bibinfo {author} {\bibfnamefont {A.~P.}\ \bibnamefont
  {Bart\'ok}}, \bibinfo {author} {\bibfnamefont {M.~C.}\ \bibnamefont {Payne}},
  \bibinfo {author} {\bibfnamefont {R.}~\bibnamefont {Kondor}}, \ and\ \bibinfo
  {author} {\bibfnamefont {G.}~\bibnamefont {Cs\'anyi}},\ }\href {\doibase
  10.1103/PhysRevLett.104.136403} {\bibfield  {journal} {\bibinfo  {journal}
  {Physical Review Letters}\ }\textbf {\bibinfo {volume} {104}},\ \bibinfo
  {pages} {136403} (\bibinfo {year} {2010})}\BibitemShut {NoStop}%
\bibitem [{\citenamefont {Rupp}\ \emph {et~al.}(2012)\citenamefont {Rupp},
  \citenamefont {Tkatchenko}, \citenamefont {M\"uller},\ and\ \citenamefont
  {von Lilienfeld}}]{Rupp2012}%
  \BibitemOpen
  \bibfield  {author} {\bibinfo {author} {\bibfnamefont {M.}~\bibnamefont
  {Rupp}}, \bibinfo {author} {\bibfnamefont {A.}~\bibnamefont {Tkatchenko}},
  \bibinfo {author} {\bibfnamefont {K.-R.}\ \bibnamefont {M\"uller}}, \ and\
  \bibinfo {author} {\bibfnamefont {O.~A.}\ \bibnamefont {von Lilienfeld}},\
  }\href {https://link.aps.org/doi/10.1103/PhysRevLett.108.058301} {\bibfield
  {journal} {\bibinfo  {journal} {Physical Review Letters}\ }\textbf {\bibinfo
  {volume} {108}},\ \bibinfo {pages} {058301} (\bibinfo {year}
  {2012})}\BibitemShut {NoStop}%
\bibitem [{\citenamefont {Thompson}\ \emph {et~al.}(2015)\citenamefont
  {Thompson}, \citenamefont {Swiler}, \citenamefont {Trott}, \citenamefont
  {Foiles},\ and\ \citenamefont {Tucker}}]{Thompson:2015dw}%
  \BibitemOpen
  \bibfield  {author} {\bibinfo {author} {\bibfnamefont {A.~P.}\ \bibnamefont
  {Thompson}}, \bibinfo {author} {\bibfnamefont {L.~P.}\ \bibnamefont
  {Swiler}}, \bibinfo {author} {\bibfnamefont {C.~R.}\ \bibnamefont {Trott}},
  \bibinfo {author} {\bibfnamefont {S.~M.}\ \bibnamefont {Foiles}}, \ and\
  \bibinfo {author} {\bibfnamefont {G.~J.}\ \bibnamefont {Tucker}},\
  }\href@noop {} {\bibfield  {journal} {\bibinfo  {journal} {Journal of
  Computational Physics}\ }\textbf {\bibinfo {volume} {285}},\ \bibinfo {pages}
  {316} (\bibinfo {year} {2015})}\BibitemShut {NoStop}%
\bibitem [{\citenamefont {Shapeev}(2016)}]{Shapeev2016}%
  \BibitemOpen
  \bibfield  {author} {\bibinfo {author} {\bibfnamefont {A.~V.}\ \bibnamefont
  {Shapeev}},\ }\href {https://epubs.siam.org/doi/abs/10.1137/15M1054183}
  {\bibfield  {journal} {\bibinfo  {journal} {Multiscale Modeling \&
  Simulation}\ }\textbf {\bibinfo {volume} {14}},\ \bibinfo {pages} {1153}
  (\bibinfo {year} {2016})}\BibitemShut {NoStop}%
\bibitem [{\citenamefont {Glielmo}\ \emph {et~al.}(2018)\citenamefont
  {Glielmo}, \citenamefont {Zeni},\ and\ \citenamefont {{De
  Vita}}}]{Glielmo2018}%
  \BibitemOpen
  \bibfield  {author} {\bibinfo {author} {\bibfnamefont {A.}~\bibnamefont
  {Glielmo}}, \bibinfo {author} {\bibfnamefont {C.}~\bibnamefont {Zeni}}, \
  and\ \bibinfo {author} {\bibfnamefont {A.}~\bibnamefont {{De Vita}}},\ }\href
  {https://journals.aps.org/prb/abstract/10.1103/PhysRevB.97.184307} {\bibfield
   {journal} {\bibinfo  {journal} {Physical Review B}\ }\textbf {\bibinfo
  {volume} {97}},\ \bibinfo {pages} {1} (\bibinfo {year} {2018})}\BibitemShut
  {NoStop}%
\bibitem [{\citenamefont {Rossi}\ and\ \citenamefont
  {Cumby}(2020)}]{Rossi2020c}%
  \BibitemOpen
  \bibfield  {author} {\bibinfo {author} {\bibfnamefont {K.}~\bibnamefont
  {Rossi}}\ and\ \bibinfo {author} {\bibfnamefont {J.}~\bibnamefont {Cumby}},\
  }\href {https://doi.org/10.1002/qua.26151} {\bibfield  {journal} {\bibinfo
  {journal} {International Journal of Quantum Chemistry}\ }\textbf {\bibinfo
  {volume} {120}} (\bibinfo {year} {2020})}\BibitemShut {NoStop}%
\bibitem [{\citenamefont {Drautz}(2019)}]{Drautz2019}%
  \BibitemOpen
  \bibfield  {author} {\bibinfo {author} {\bibfnamefont {R.}~\bibnamefont
  {Drautz}},\ }\href {\doibase 10.1103/PhysRevB.99.014104} {\bibfield
  {journal} {\bibinfo  {journal} {Phys. Rev. B}\ }\textbf {\bibinfo {volume}
  {99}},\ \bibinfo {pages} {014104} (\bibinfo {year} {2019})}\BibitemShut
  {NoStop}%
\bibitem [{\citenamefont {Bachmayr}\ \emph {et~al.}(2020)\citenamefont
  {Bachmayr}, \citenamefont {Csanyi}, \citenamefont {Drautz}, \citenamefont
  {Dusson}, \citenamefont {Etter}, \citenamefont {van~der Oord},\ and\
  \citenamefont {Ortner}}]{Bachmayr2020}%
  \BibitemOpen
  \bibfield  {author} {\bibinfo {author} {\bibfnamefont {M.}~\bibnamefont
  {Bachmayr}}, \bibinfo {author} {\bibfnamefont {G.}~\bibnamefont {Csanyi}},
  \bibinfo {author} {\bibfnamefont {R.}~\bibnamefont {Drautz}}, \bibinfo
  {author} {\bibfnamefont {G.}~\bibnamefont {Dusson}}, \bibinfo {author}
  {\bibfnamefont {S.}~\bibnamefont {Etter}}, \bibinfo {author} {\bibfnamefont
  {C.}~\bibnamefont {van~der Oord}}, \ and\ \bibinfo {author} {\bibfnamefont
  {C.}~\bibnamefont {Ortner}},\ }\href@noop {} {\enquote {\bibinfo {title}
  {Atomic cluster expansion: Completeness, efficiency and stability},}\ }
  (\bibinfo {year} {2020}),\ \Eprint {http://arxiv.org/abs/1911.03550}
  {arXiv:1911.03550 [math.NA]} \BibitemShut {NoStop}%
\bibitem [{\citenamefont {Drautz}(2020)}]{Drautz2020}%
  \BibitemOpen
  \bibfield  {author} {\bibinfo {author} {\bibfnamefont {R.}~\bibnamefont
  {Drautz}},\ }\href {\doibase 10.1103/PhysRevB.102.024104} {\bibfield
  {journal} {\bibinfo  {journal} {Phys. Rev. B}\ }\textbf {\bibinfo {volume}
  {102}},\ \bibinfo {pages} {024104} (\bibinfo {year} {2020})}\BibitemShut
  {NoStop}%
\bibitem [{\citenamefont {Kocer}\ \emph {et~al.}(2019)\citenamefont {Kocer},
  \citenamefont {Mason},\ and\ \citenamefont {Erturk}}]{Kocer2019}%
  \BibitemOpen
  \bibfield  {author} {\bibinfo {author} {\bibfnamefont {E.}~\bibnamefont
  {Kocer}}, \bibinfo {author} {\bibfnamefont {J.~K.}\ \bibnamefont {Mason}}, \
  and\ \bibinfo {author} {\bibfnamefont {H.}~\bibnamefont {Erturk}},\ }\href
  {https://doi.org/10.1063/1.5086167} {\bibfield  {journal} {\bibinfo
  {journal} {The Journal of Chemical Physics}\ }\textbf {\bibinfo {volume}
  {150}},\ \bibinfo {pages} {154102} (\bibinfo {year} {2019})}\BibitemShut
  {NoStop}%
\bibitem [{\citenamefont {Glielmo}\ \emph {et~al.}(2019)\citenamefont
  {Glielmo}, \citenamefont {Zeni}, \citenamefont {Fekete},\ and\ \citenamefont
  {De~Vita}}]{Glielmo2019}%
  \BibitemOpen
  \bibfield  {author} {\bibinfo {author} {\bibfnamefont {A.}~\bibnamefont
  {Glielmo}}, \bibinfo {author} {\bibfnamefont {C.}~\bibnamefont {Zeni}},
  \bibinfo {author} {\bibfnamefont {{\'{A}}.}~\bibnamefont {Fekete}}, \ and\
  \bibinfo {author} {\bibfnamefont {A.}~\bibnamefont {De~Vita}},\ }\href
  {https://arxiv.org/abs/1905.07626} {\bibfield  {journal} {\bibinfo  {journal}
  {arXiv:1905.07626}\ } (\bibinfo {year} {2019})},\ \Eprint
  {http://arxiv.org/abs/1905.07626} {arXiv:1905.07626 [Physics - Computational
  Physics]} \BibitemShut {NoStop}%
\bibitem [{\citenamefont {Blum}\ and\ \citenamefont
  {Reymond}(2009)}]{Blum2009}%
  \BibitemOpen
  \bibfield  {author} {\bibinfo {author} {\bibfnamefont {L.~C.}\ \bibnamefont
  {Blum}}\ and\ \bibinfo {author} {\bibfnamefont {J.-L.}\ \bibnamefont
  {Reymond}},\ }\href {https://doi.org/10.1021/ja902302h} {\bibfield  {journal}
  {\bibinfo  {journal} {Journal of the American Chemical Society}\ }\textbf
  {\bibinfo {volume} {131}},\ \bibinfo {pages} {8732} (\bibinfo {year}
  {2009})}\BibitemShut {NoStop}%
\bibitem [{\citenamefont {Zuo}\ \emph {et~al.}(2020)\citenamefont {Zuo},
  \citenamefont {Chen}, \citenamefont {Li}, \citenamefont {Deng}, \citenamefont
  {Chen}, \citenamefont {Behler}, \citenamefont {Cs{\'a}nyi}, \citenamefont
  {Shapeev}, \citenamefont {Thompson}, \citenamefont {Wood} \emph
  {et~al.}}]{zuo2020performance}%
  \BibitemOpen
  \bibfield  {author} {\bibinfo {author} {\bibfnamefont {Y.}~\bibnamefont
  {Zuo}}, \bibinfo {author} {\bibfnamefont {C.}~\bibnamefont {Chen}}, \bibinfo
  {author} {\bibfnamefont {X.}~\bibnamefont {Li}}, \bibinfo {author}
  {\bibfnamefont {Z.}~\bibnamefont {Deng}}, \bibinfo {author} {\bibfnamefont
  {Y.}~\bibnamefont {Chen}}, \bibinfo {author} {\bibfnamefont {J.}~\bibnamefont
  {Behler}}, \bibinfo {author} {\bibfnamefont {G.}~\bibnamefont {Cs{\'a}nyi}},
  \bibinfo {author} {\bibfnamefont {A.~V.}\ \bibnamefont {Shapeev}}, \bibinfo
  {author} {\bibfnamefont {A.~P.}\ \bibnamefont {Thompson}}, \bibinfo {author}
  {\bibfnamefont {M.~A.}\ \bibnamefont {Wood}},  \emph {et~al.},\ }\href
  {\doibase doi: 10.1021/acs.jpca.9b08723} {\bibfield  {journal} {\bibinfo
  {journal} {The Journal of Physical Chemistry A}\ }\textbf {\bibinfo {volume}
  {124}},\ \bibinfo {pages} {731} (\bibinfo {year} {2020})}\BibitemShut
  {NoStop}%
\bibitem [{\citenamefont {Kocer}\ \emph {et~al.}(2020)\citenamefont {Kocer},
  \citenamefont {Mason},\ and\ \citenamefont {Erturk}}]{Kocer2020}%
  \BibitemOpen
  \bibfield  {author} {\bibinfo {author} {\bibfnamefont {E.}~\bibnamefont
  {Kocer}}, \bibinfo {author} {\bibfnamefont {J.~K.}\ \bibnamefont {Mason}}, \
  and\ \bibinfo {author} {\bibfnamefont {H.}~\bibnamefont {Erturk}},\ }\href
  {https://doi.org/10.1063/1.5111045} {\bibfield  {journal} {\bibinfo
  {journal} {AIP Advances}\ }\textbf {\bibinfo {volume} {10}},\ \bibinfo
  {pages} {015021} (\bibinfo {year} {2020})}\BibitemShut {NoStop}%
\bibitem [{\citenamefont {Glielmo}\ \emph {et~al.}(2017)\citenamefont
  {Glielmo}, \citenamefont {Sollich},\ and\ \citenamefont {{De
  Vita}}}]{Glielmo2017}%
  \BibitemOpen
  \bibfield  {author} {\bibinfo {author} {\bibfnamefont {A.}~\bibnamefont
  {Glielmo}}, \bibinfo {author} {\bibfnamefont {P.}~\bibnamefont {Sollich}}, \
  and\ \bibinfo {author} {\bibfnamefont {A.}~\bibnamefont {{De Vita}}},\ }\href
  {https://journals.aps.org/prb/abstract/10.1103/PhysRevB.95.214302} {\bibfield
   {journal} {\bibinfo  {journal} {Physical Review B}\ }\textbf {\bibinfo
  {volume} {95}},\ \bibinfo {pages} {1} (\bibinfo {year} {2017})}\BibitemShut
  {NoStop}%
\bibitem [{\citenamefont {Artrith}\ \emph {et~al.}(2017)\citenamefont
  {Artrith}, \citenamefont {Urban},\ and\ \citenamefont {Ceder}}]{Artrith2017}%
  \BibitemOpen
  \bibfield  {author} {\bibinfo {author} {\bibfnamefont {N.}~\bibnamefont
  {Artrith}}, \bibinfo {author} {\bibfnamefont {A.}~\bibnamefont {Urban}}, \
  and\ \bibinfo {author} {\bibfnamefont {G.}~\bibnamefont {Ceder}},\ }\href
  {https://link.aps.org/doi/10.1103/PhysRevB.96.014112} {\bibfield  {journal}
  {\bibinfo  {journal} {Physical Review B}\ }\textbf {\bibinfo {volume} {96}},\
  \bibinfo {pages} {014112} (\bibinfo {year} {2017})}\BibitemShut {NoStop}%
\bibitem [{\citenamefont {Ramakrishnan}\ \emph {et~al.}(2014)\citenamefont
  {Ramakrishnan}, \citenamefont {Dral}, \citenamefont {Rupp},\ and\
  \citenamefont {Anatole~von Lilienfeld}}]{Ramakrishnan_2014}%
  \BibitemOpen
  \bibfield  {author} {\bibinfo {author} {\bibfnamefont {R.}~\bibnamefont
  {Ramakrishnan}}, \bibinfo {author} {\bibfnamefont {P.}~\bibnamefont {Dral}},
  \bibinfo {author} {\bibfnamefont {M.}~\bibnamefont {Rupp}}, \ and\ \bibinfo
  {author} {\bibfnamefont {O.}~\bibnamefont {Anatole~von Lilienfeld}},\ }\href
  {\doibase 10.6084/m9.figshare.c.978904.v5} {\enquote {\bibinfo {title}
  {Quantum chemistry structures and properties of 134 kilo molecules},}\ }
  (\bibinfo {year} {2014})\BibitemShut {NoStop}%
\bibitem [{\citenamefont {Unke}\ and\ \citenamefont {Meuwly}(2018)}]{Unke2018}%
  \BibitemOpen
  \bibfield  {author} {\bibinfo {author} {\bibfnamefont {O.~T.}\ \bibnamefont
  {Unke}}\ and\ \bibinfo {author} {\bibfnamefont {M.}~\bibnamefont {Meuwly}},\
  }\href@noop {} {\bibfield  {journal} {\bibinfo  {journal} {Journal of
  Chemical Physics}\ }\textbf {\bibinfo {volume} {148}} (\bibinfo {year}
  {2018})}\BibitemShut {NoStop}%
\bibitem [{\citenamefont {Willatt}\ \emph {et~al.}(2019)\citenamefont
  {Willatt}, \citenamefont {Musil},\ and\ \citenamefont
  {Ceriotti}}]{Willatt2019}%
  \BibitemOpen
  \bibfield  {author} {\bibinfo {author} {\bibfnamefont {M.~J.}\ \bibnamefont
  {Willatt}}, \bibinfo {author} {\bibfnamefont {F.}~\bibnamefont {Musil}}, \
  and\ \bibinfo {author} {\bibfnamefont {M.}~\bibnamefont {Ceriotti}},\ }\href
  {https://aip.scitation.org/doi/full/10.1063/1.5090481} {\bibfield  {journal}
  {\bibinfo  {journal} {Journal of Chemical Physics}\ }\textbf {\bibinfo
  {volume} {154110}} (\bibinfo {year} {2019})}\BibitemShut {NoStop}%
\bibitem [{\citenamefont {Jinnouchi}\ \emph {et~al.}(2020)\citenamefont
  {Jinnouchi}, \citenamefont {Karsai}, \citenamefont {Verdi}, \citenamefont
  {Asahi},\ and\ \citenamefont {Kresse}}]{Jinnouchi2020}%
  \BibitemOpen
  \bibfield  {author} {\bibinfo {author} {\bibfnamefont {R.}~\bibnamefont
  {Jinnouchi}}, \bibinfo {author} {\bibfnamefont {F.}~\bibnamefont {Karsai}},
  \bibinfo {author} {\bibfnamefont {C.}~\bibnamefont {Verdi}}, \bibinfo
  {author} {\bibfnamefont {R.}~\bibnamefont {Asahi}}, \ and\ \bibinfo {author}
  {\bibfnamefont {G.}~\bibnamefont {Kresse}},\ }\href {\doibase
  10.1063/5.0009491} {\bibfield  {journal} {\bibinfo  {journal} {Journal of
  Chemical Physics}\ }\textbf {\bibinfo {volume} {152}},\ \bibinfo {pages}
  {234102} (\bibinfo {year} {2020})}\BibitemShut {NoStop}%
\bibitem [{\citenamefont {Lysogorskiy}\ \emph {et~al.}(2021)\citenamefont
  {Lysogorskiy}, \citenamefont {van~der Oord}, \citenamefont {Bochkarev},
  \citenamefont {Menon}, \citenamefont {Rinaldi}, \citenamefont
  {Hammerschmidt}, \citenamefont {Mrovec}, \citenamefont {Thompson},
  \citenamefont {Cs{\'a}nyi}, \citenamefont {Ortner} \emph
  {et~al.}}]{lysogorskiy2021performant}%
  \BibitemOpen
  \bibfield  {author} {\bibinfo {author} {\bibfnamefont {Y.}~\bibnamefont
  {Lysogorskiy}}, \bibinfo {author} {\bibfnamefont {C.}~\bibnamefont {van~der
  Oord}}, \bibinfo {author} {\bibfnamefont {A.}~\bibnamefont {Bochkarev}},
  \bibinfo {author} {\bibfnamefont {S.}~\bibnamefont {Menon}}, \bibinfo
  {author} {\bibfnamefont {M.}~\bibnamefont {Rinaldi}}, \bibinfo {author}
  {\bibfnamefont {T.}~\bibnamefont {Hammerschmidt}}, \bibinfo {author}
  {\bibfnamefont {M.}~\bibnamefont {Mrovec}}, \bibinfo {author} {\bibfnamefont
  {A.}~\bibnamefont {Thompson}}, \bibinfo {author} {\bibfnamefont
  {G.}~\bibnamefont {Cs{\'a}nyi}}, \bibinfo {author} {\bibfnamefont
  {C.}~\bibnamefont {Ortner}},  \emph {et~al.},\ }\href
  {https://arxiv.org/abs/2103.00814} {\bibfield  {journal} {\bibinfo  {journal}
  {arXiv preprint arXiv:2103.00814}\ } (\bibinfo {year} {2021})}\BibitemShut
  {NoStop}%
\bibitem [{\citenamefont {Glielmo}\ \emph {et~al.}(2021)\citenamefont
  {Glielmo}, \citenamefont {Zeni}, \citenamefont {Cheng}, \citenamefont
  {Csanyi},\ and\ \citenamefont {Laio}}]{glielmo2021ranking}%
  \BibitemOpen
  \bibfield  {author} {\bibinfo {author} {\bibfnamefont {A.}~\bibnamefont
  {Glielmo}}, \bibinfo {author} {\bibfnamefont {C.}~\bibnamefont {Zeni}},
  \bibinfo {author} {\bibfnamefont {B.}~\bibnamefont {Cheng}}, \bibinfo
  {author} {\bibfnamefont {G.}~\bibnamefont {Csanyi}}, \ and\ \bibinfo {author}
  {\bibfnamefont {A.}~\bibnamefont {Laio}},\ }\href@noop {} {\enquote {\bibinfo
  {title} {Ranking the information content of distance measures},}\ } (\bibinfo
  {year} {2021}),\ \Eprint {http://arxiv.org/abs/2104.15079} {arXiv:2104.15079
  [stat.ML]} \BibitemShut {NoStop}%
\bibitem [{\citenamefont {Pearson}(1901)}]{Pearson1901}%
  \BibitemOpen
  \bibfield  {author} {\bibinfo {author} {\bibfnamefont {K.}~\bibnamefont
  {Pearson}},\ }\href {\doibase 10.1080/14786440109462720} {\bibfield
  {journal} {\bibinfo  {journal} {The London, Edinburgh, and Dublin
  Philosophical Magazine and Journal of Science}\ }\textbf {\bibinfo {volume}
  {2}},\ \bibinfo {pages} {559} (\bibinfo {year} {1901})}\BibitemShut {NoStop}%
\bibitem [{\citenamefont {Cubuk}\ \emph {et~al.}(2017)\citenamefont {Cubuk},
  \citenamefont {Malone}, \citenamefont {Onat}, \citenamefont {Waterland},\
  and\ \citenamefont {Kaxiras}}]{cubuk2017representations}%
  \BibitemOpen
  \bibfield  {author} {\bibinfo {author} {\bibfnamefont {E.~D.}\ \bibnamefont
  {Cubuk}}, \bibinfo {author} {\bibfnamefont {B.~D.}\ \bibnamefont {Malone}},
  \bibinfo {author} {\bibfnamefont {B.}~\bibnamefont {Onat}}, \bibinfo {author}
  {\bibfnamefont {A.}~\bibnamefont {Waterland}}, \ and\ \bibinfo {author}
  {\bibfnamefont {E.}~\bibnamefont {Kaxiras}},\ }\href {\doibase
  10.1063/1.4990503} {\bibfield  {journal} {\bibinfo  {journal} {The Journal of
  chemical physics}\ }\textbf {\bibinfo {volume} {147}},\ \bibinfo {pages}
  {024104} (\bibinfo {year} {2017})},\ \Eprint
  {http://arxiv.org/abs/https://doi.org/10.1063/1.4990503}
  {https://doi.org/10.1063/1.4990503} \BibitemShut {NoStop}%
\bibitem [{\citenamefont {Efron}\ \emph {et~al.}(2004)\citenamefont {Efron},
  \citenamefont {Hastie}, \citenamefont {Johnstone}, \citenamefont {Tibshirani}
  \emph {et~al.}}]{efron2004least}%
  \BibitemOpen
  \bibfield  {author} {\bibinfo {author} {\bibfnamefont {B.}~\bibnamefont
  {Efron}}, \bibinfo {author} {\bibfnamefont {T.}~\bibnamefont {Hastie}},
  \bibinfo {author} {\bibfnamefont {I.}~\bibnamefont {Johnstone}}, \bibinfo
  {author} {\bibfnamefont {R.}~\bibnamefont {Tibshirani}},  \emph {et~al.},\
  }\href {\doibase 10.1214/009053604000000067} {\bibfield  {journal} {\bibinfo
  {journal} {Annals of statistics}\ }\textbf {\bibinfo {volume} {32}},\
  \bibinfo {pages} {407} (\bibinfo {year} {2004})}\BibitemShut {NoStop}%
\bibitem [{\citenamefont {Benoit}\ \emph {et~al.}(2021)\citenamefont {Benoit},
  \citenamefont {Amodeo}, \citenamefont {Combettes}, \citenamefont {Khaled},
  \citenamefont {Roux},\ and\ \citenamefont {Lam}}]{Benoit2021}%
  \BibitemOpen
  \bibfield  {author} {\bibinfo {author} {\bibfnamefont {M.}~\bibnamefont
  {Benoit}}, \bibinfo {author} {\bibfnamefont {J.}~\bibnamefont {Amodeo}},
  \bibinfo {author} {\bibfnamefont {S.}~\bibnamefont {Combettes}}, \bibinfo
  {author} {\bibfnamefont {I.}~\bibnamefont {Khaled}}, \bibinfo {author}
  {\bibfnamefont {A.}~\bibnamefont {Roux}}, \ and\ \bibinfo {author}
  {\bibfnamefont {J.}~\bibnamefont {Lam}},\ }\href {\doibase
  10.1088/2632-2153/abc9fd} {\bibfield  {journal} {\bibinfo  {journal} {Machine
  Learning: Science and Technology}\ } (\bibinfo {year} {2021}),\
  10.1088/2632-2153/abc9fd},\ \Eprint {http://arxiv.org/abs/1912.10761}
  {arXiv:1912.10761} \BibitemShut {NoStop}%
\bibitem [{\citenamefont {Ghiringhelli}\ \emph {et~al.}(2015)\citenamefont
  {Ghiringhelli}, \citenamefont {Vybiral}, \citenamefont {Levchenko},
  \citenamefont {Draxl},\ and\ \citenamefont
  {Scheffler}}]{ghiringhelli2015big}%
  \BibitemOpen
  \bibfield  {author} {\bibinfo {author} {\bibfnamefont {L.~M.}\ \bibnamefont
  {Ghiringhelli}}, \bibinfo {author} {\bibfnamefont {J.}~\bibnamefont
  {Vybiral}}, \bibinfo {author} {\bibfnamefont {S.~V.}\ \bibnamefont
  {Levchenko}}, \bibinfo {author} {\bibfnamefont {C.}~\bibnamefont {Draxl}}, \
  and\ \bibinfo {author} {\bibfnamefont {M.}~\bibnamefont {Scheffler}},\
  }\href@noop {} {\bibfield  {journal} {\bibinfo  {journal} {Physical review
  letters}\ }\textbf {\bibinfo {volume} {114}},\ \bibinfo {pages} {105503}
  (\bibinfo {year} {2015})}\BibitemShut {NoStop}%
\bibitem [{\citenamefont {Ouyang}\ \emph {et~al.}(2018)\citenamefont {Ouyang},
  \citenamefont {Curtarolo}, \citenamefont {Ahmetcik}, \citenamefont
  {Scheffler},\ and\ \citenamefont {Ghiringhelli}}]{ouyang2018sisso}%
  \BibitemOpen
  \bibfield  {author} {\bibinfo {author} {\bibfnamefont {R.}~\bibnamefont
  {Ouyang}}, \bibinfo {author} {\bibfnamefont {S.}~\bibnamefont {Curtarolo}},
  \bibinfo {author} {\bibfnamefont {E.}~\bibnamefont {Ahmetcik}}, \bibinfo
  {author} {\bibfnamefont {M.}~\bibnamefont {Scheffler}}, \ and\ \bibinfo
  {author} {\bibfnamefont {L.~M.}\ \bibnamefont {Ghiringhelli}},\ }\href
  {\doibase 10.1103/PhysRevMaterials.2.083802} {\bibfield  {journal} {\bibinfo
  {journal} {Physical Review Materials}\ }\textbf {\bibinfo {volume} {2}},\
  \bibinfo {pages} {083802} (\bibinfo {year} {2018})}\BibitemShut {NoStop}%
\bibitem [{\citenamefont {Nelson}\ \emph {et~al.}(2013)\citenamefont {Nelson},
  \citenamefont {Hart}, \citenamefont {Zhou}, \citenamefont
  {Ozoli{\c{n}}{\v{s}}} \emph {et~al.}}]{nelson2013compressive}%
  \BibitemOpen
  \bibfield  {author} {\bibinfo {author} {\bibfnamefont {L.~J.}\ \bibnamefont
  {Nelson}}, \bibinfo {author} {\bibfnamefont {G.~L.}\ \bibnamefont {Hart}},
  \bibinfo {author} {\bibfnamefont {F.}~\bibnamefont {Zhou}}, \bibinfo {author}
  {\bibfnamefont {V.}~\bibnamefont {Ozoli{\c{n}}{\v{s}}}},  \emph {et~al.},\
  }\href {\doibase 10.1103/PhysRevB.87.035125} {\bibfield  {journal} {\bibinfo
  {journal} {Physical Review B}\ }\textbf {\bibinfo {volume} {87}},\ \bibinfo
  {pages} {035125} (\bibinfo {year} {2013})}\BibitemShut {NoStop}%
\bibitem [{\citenamefont {Basilevsky}(2013)}]{basilevsky2013applied}%
  \BibitemOpen
  \bibfield  {author} {\bibinfo {author} {\bibfnamefont {A.}~\bibnamefont
  {Basilevsky}},\ }\href@noop {} {\emph {\bibinfo {title} {Applied matrix
  algebra in the statistical sciences}}}\ (\bibinfo  {publisher} {Courier
  Corporation},\ \bibinfo {year} {2013})\BibitemShut {NoStop}%
\end{thebibliography}%



\appendix
\clearpage
\newpage

\renewcommand{\thetable}{S\arabic{table}}
\renewcommand{\thefigure}{S\arabic{figure}}

\setcounter{figure}{0}

\section{Supplementary material}
\subsection{Approximation of a delta function}
To better probe the intrinsic differences between the three radial basis function sets, we look at how well these radial basis function sets approximate a Dirac delta function as $n_{MAX}$ increases.
To do so, we must first modify the radial basis functions $g_n(x)$ to obtain orthogonal basis sets for functions defined in the $\left[0, 1 \right]$ interval.
Therefore, we define the set of functions $\widetilde{g}_n(x)$ as follows for the SSB basis set:
\begin{equation}
\widetilde{g}_n(x) = \dfrac{g_n(x)}{x}, 
\label{eq:bessel_delta}
\end{equation}
and for the SC, and NSC radial basis sets:
\begin{equation}
\widetilde{g}_n(x) = \dfrac{g_n(x)}{\cos(x)-1}.
\label{eq:chebyshev_delta}
\end{equation}
We can now compute the approximation of a Dirac delta function placed at $x^*=0.5$ as the sum over the modified radial basis functions $\widetilde{g}_n(x)$ multiplied by the value these same functions have at $x=x^*$.
\\

\begin{figure}[h!]
    \centering
    \includegraphics[width=6.cm]{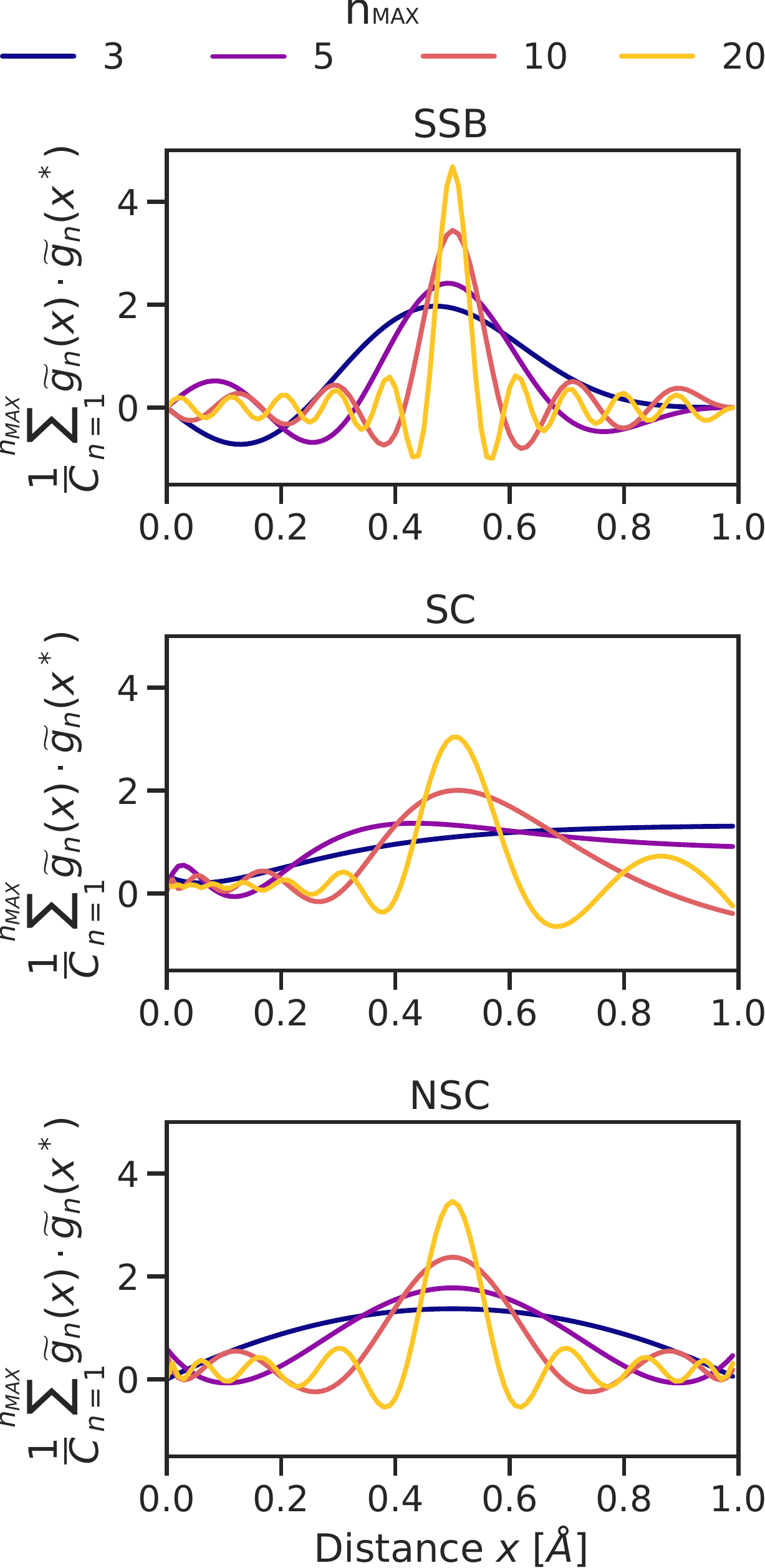}
    \caption{Approximation of a Dirac delta function placed at $x^* = 0.5$ by  (top to bottom) SSB, SC, and NSC radial basis sets, using varying numbers of radial basis $n_{MAX}$ (in color).}
    \label{fig:delta_approx}
\end{figure}

In Figure \ref{fig:delta_approx}, we display the value of $\dfrac{1}{C} \sum_{n=1}^{n_{MAX}} \widetilde{g}_n(x) \cdot \widetilde{g}_n(x^*)$ for the SSB, SC, and NSC radial basis sets, for $n_{MAX}$ = 3, 5, 10, and 20, and for $x^* = 0.5$.
$C$ is here a normalization factor equal to the square root of the square integral of $\sum_{n=1}^{n_{MAX}} \widetilde{g}_n(x) \cdot \widetilde{g}_n(x^*)$ from 0 to 1.
These profiles can be considered to be approximations of delta functions centered at $x^*$.
We observe in Figure \ref{fig:delta_approx} how the SSB radial basis set better approximates the Dirac delta placed at $x^*$ = 0.5 w.r.t. the SC and NSC basis sets.
This is especially evident for $n_{MAX} = 3$ (dark blue line), where only the SSB radial basis has a sharp maximum around $x^* = 0.5$.
\\

Since atoms are represented by delta functions in the real space within the ACE framework, the fact that SSB radial basis sets are good approximators of Dirac delta functions, even at low values of $n_{MAX}$, gives some insight relative to the better performance of this particular basis set w.r.t. the others when $n_{MAX}$ is small.
\clearpage
\subsection{Balancing angular and radial basis number}
The computational cost of evaluating the descriptor scales both as $\mathcal{O}(n_{MAX} \cdot l_{MAX}^2)$ for the computation of the $c_{nlm}$ factors, and as $\mathcal{O}(n_{MAX}^2 \cdot l_{MAX})$ for the computation of the descriptor $\mathbf{q}$.
Therefore, we look at the role played by the balance between the number of angular and radial basis employed on the prediction accuracy, trying to strike an optimal balance  between $n_{MAX}$ and $l_{MAX}$ to increase prediction accuracy.
\begin{figure}[h!]
    \centering
    \includegraphics[width=6.8cm]{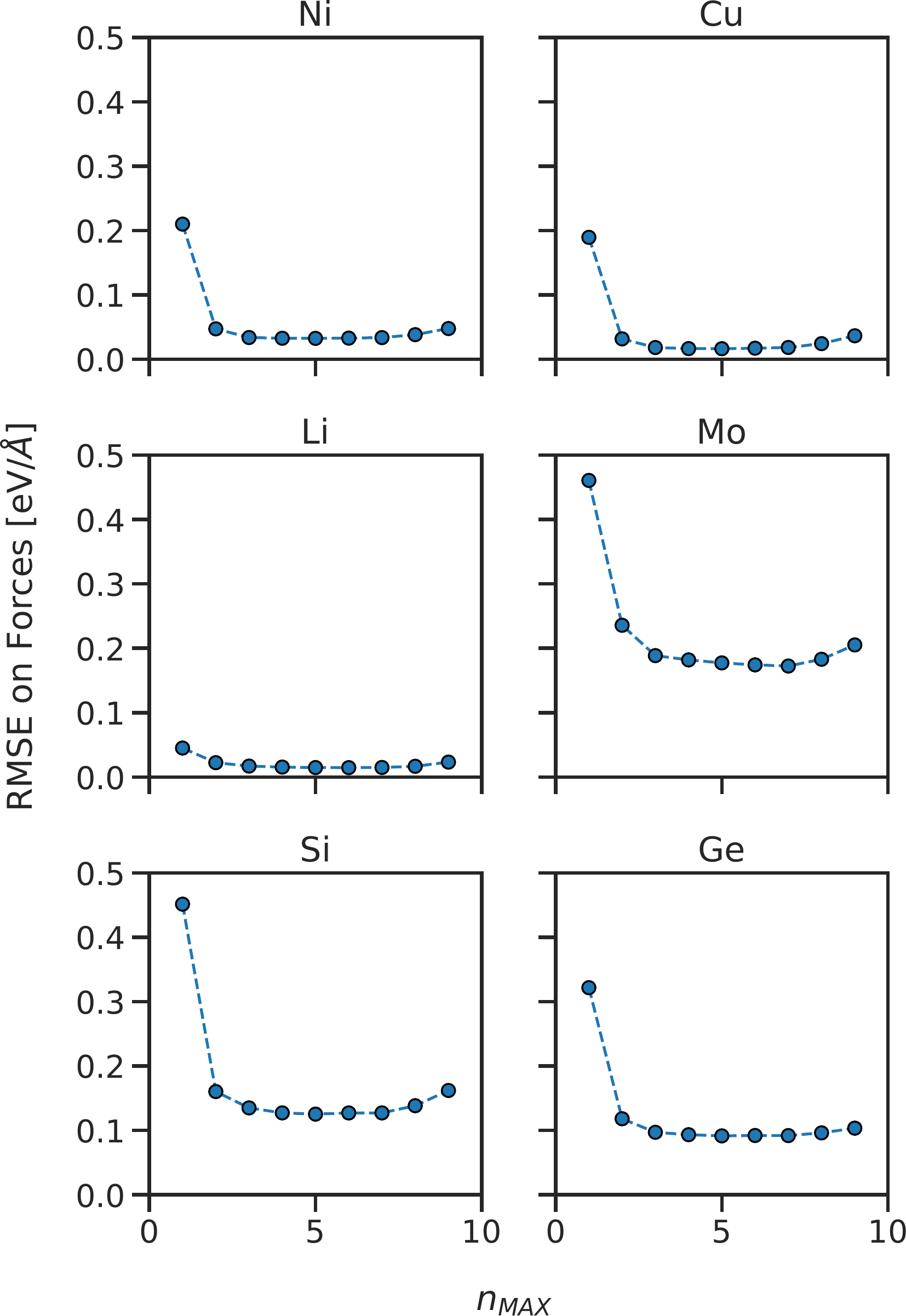}
\caption{RMSE on atomic forces as a function of the number of SSB radial basis functions employed ($n_{MAX}$) when the sum $n_{MAX}$+$l_{MAX}$ is kept constant at 12.}
\label{fig:si_constant_sum_force_zuo}
\end{figure}
\begin{figure}[h!]
    \centering
    \includegraphics[width=6.8cm]{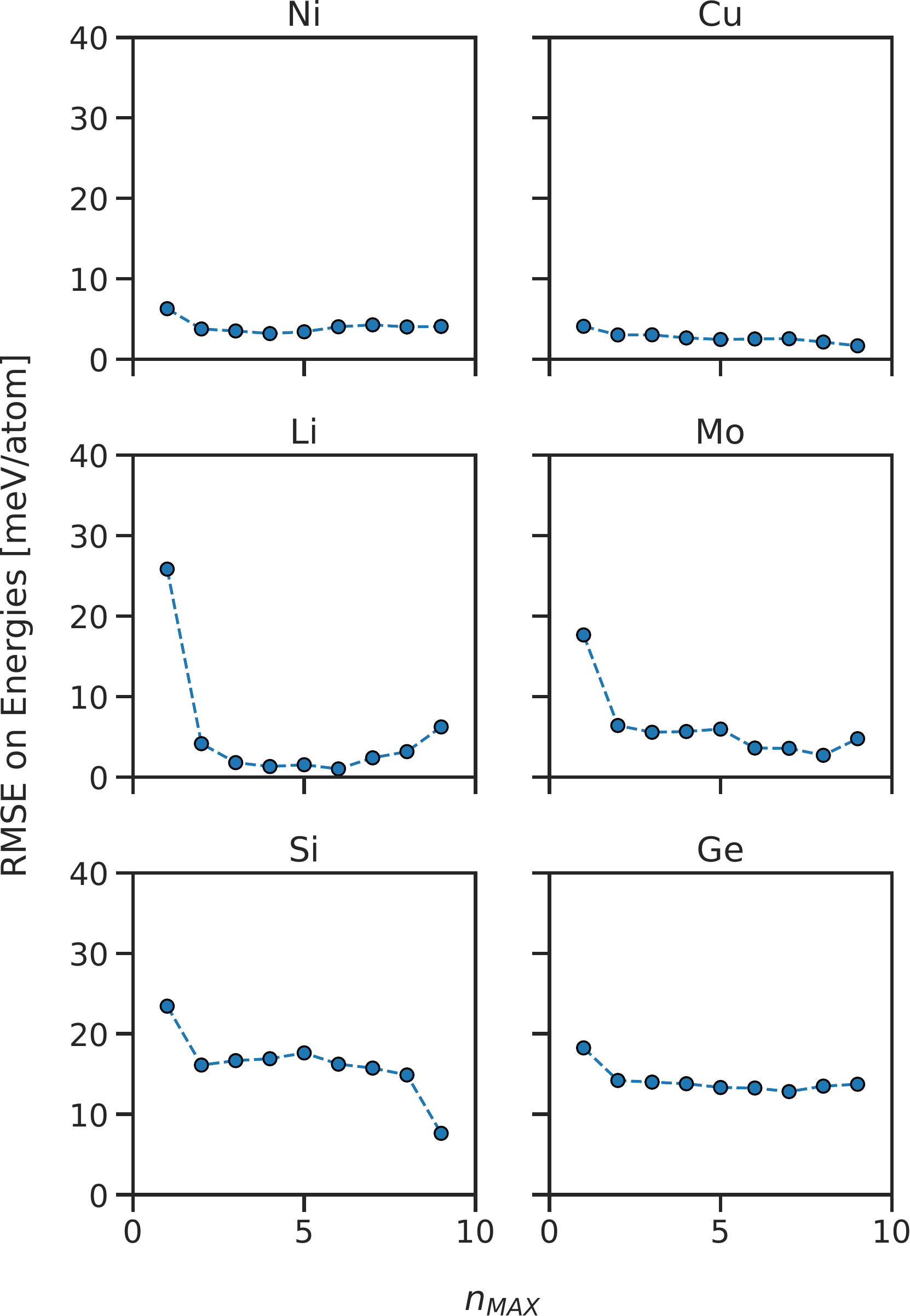}
\caption{RMSE on atomic energy as a function of the number of SSB radial basis functions employed ($n_{MAX}$) when the sum $ns$+$ls$ is kept constant at 12.}
\label{fig:si_constant_sum_energy_zuo}
\end{figure}
 For the case of the six single-element datasets introduced in Ref.~\onlinecite{zuo2020performance}, we display the validation accuracy on forces and energies in a setup similar to the one of Section III B, where the sum $n_{MAX}$ + $l_{MAX}$ is kept equal to 12.
Fig.~\ref{fig:si_constant_sum_force_zuo} illustrates how
 descriptors where $n_{MAX}$ is comparable to $l_{MAX}$ incur the lowest error on forces. 
 This trend is mostly present also in Fig.~\ref{fig:si_constant_sum_energy_zuo}, except for the case of Si, where the RMSE on energies is sensibly reduced for $n_{MAX} = 9$.
\clearpage

\subsection{PCA and LASSO dimensionality reduction for SC basis functions}
Here we report graphs that mirror Figures 5, 6, 7, and 8 of the main text, but for the case of SC radial basis functions.
\begin{figure}[h!]
    \includegraphics[width=7.5cm]{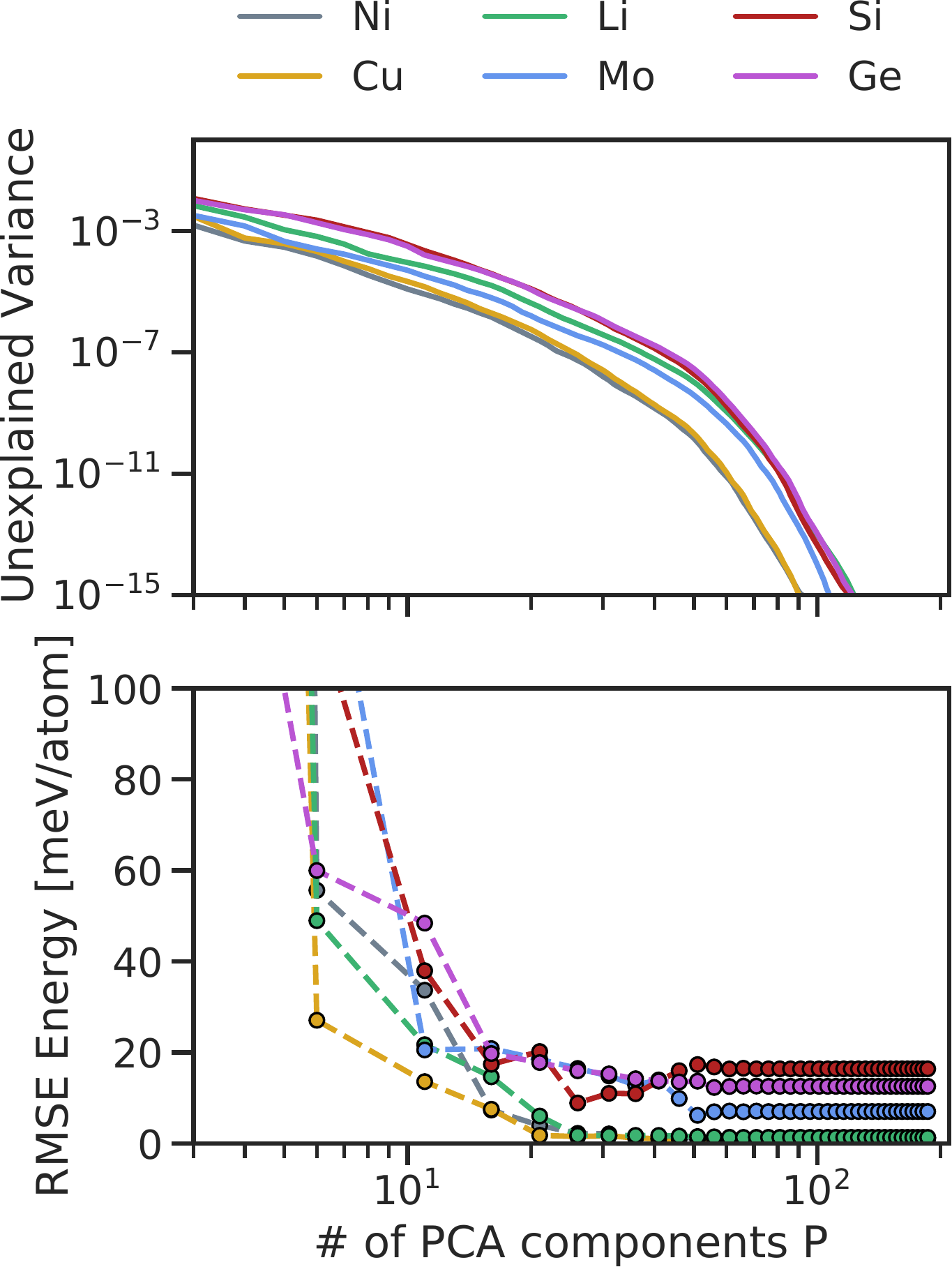}
     \caption{Top panel: data unexplained variance as a function of the number of PCA components accounted for.
     Bottom panel: RMSE on energies incurred by RR potentials employing the reduced descriptor $\mathbf{Q}^{PCA}_{P}$ on the validation set, as a function of the number of PCA components $P$.}
    \label{fig:pca_sc}
\end{figure}
\begin{figure}[h!]
    \includegraphics[width=6.5cm]{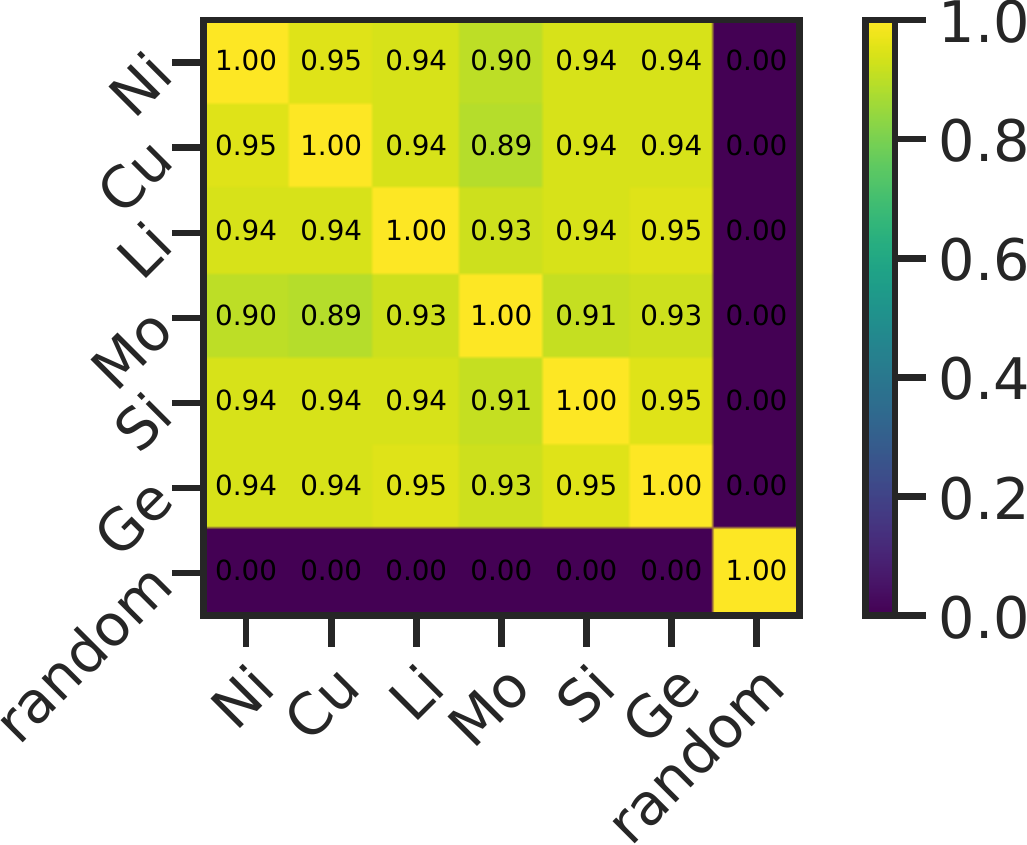}
    \caption{Heatmap displaying the fraction of dimensions shared by the sub-spaces generated by the first 80 PCA-selected directions of the descriptors $\mathbf{Q}$ among couples of single-element datasets.
    The random label indicates a sub-space generated by taking 80 random orthogonal vectors in the space of the $\mathbf{Q}$ vectors.}
    \label{fig:pca_shared_dimensions_sc}
\end{figure}
\begin{figure}[h!]
    \includegraphics[width=7.5cm]{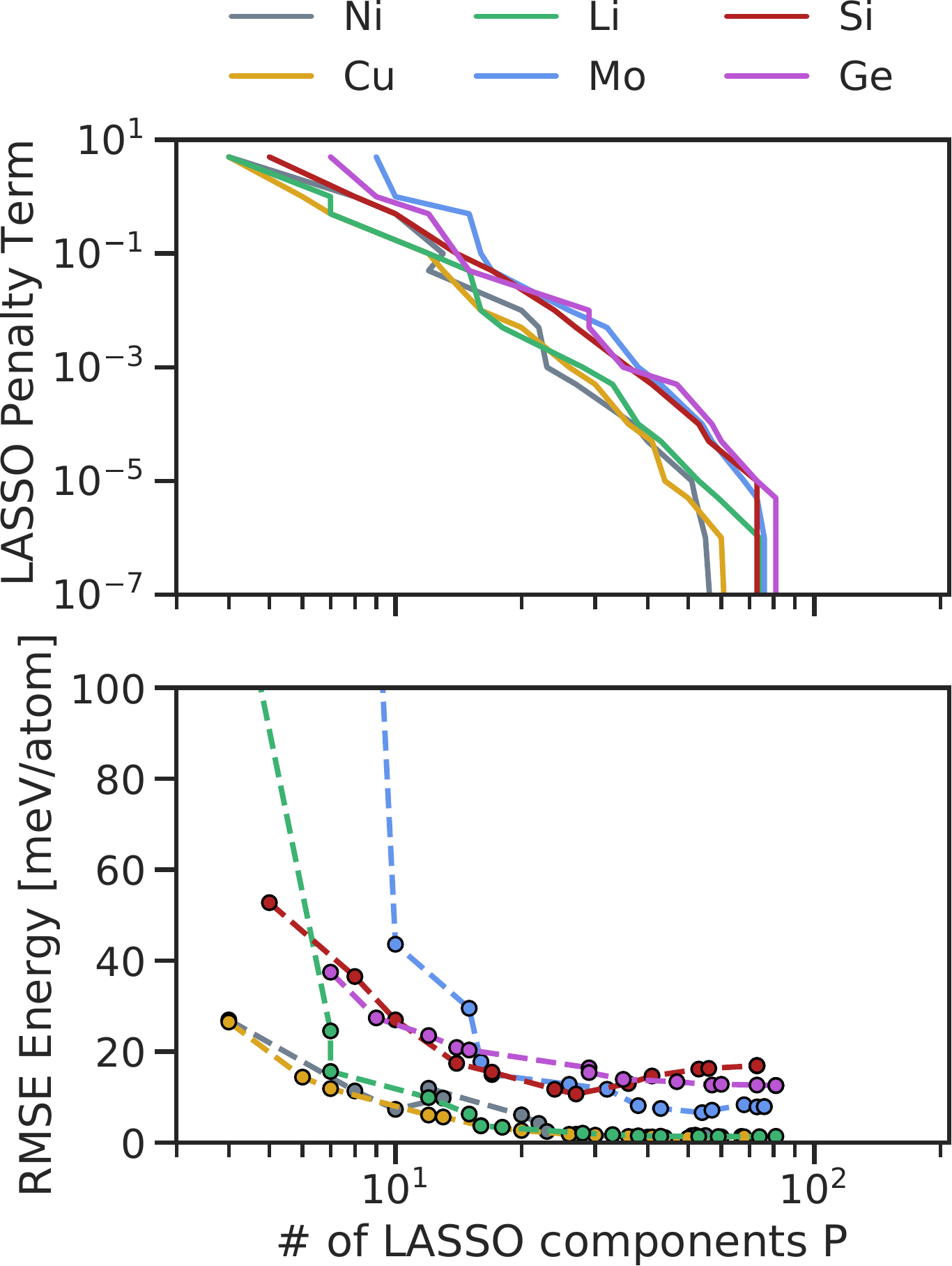}
     \caption{Top panel: LASSO penalty term as a function of the number of LASSO components accounted for.
     Bottom panel: RMSE on energies incurred by RR potentials employing the reduced descriptor $\mathbf{Q}^{LASSO}_{P}$ on the validation set, as a function of the number of LASSO components $P$.}
    \label{fig:lasso_sc}
\end{figure}
\begin{figure}[h!]
    \includegraphics[width=6.5cm]{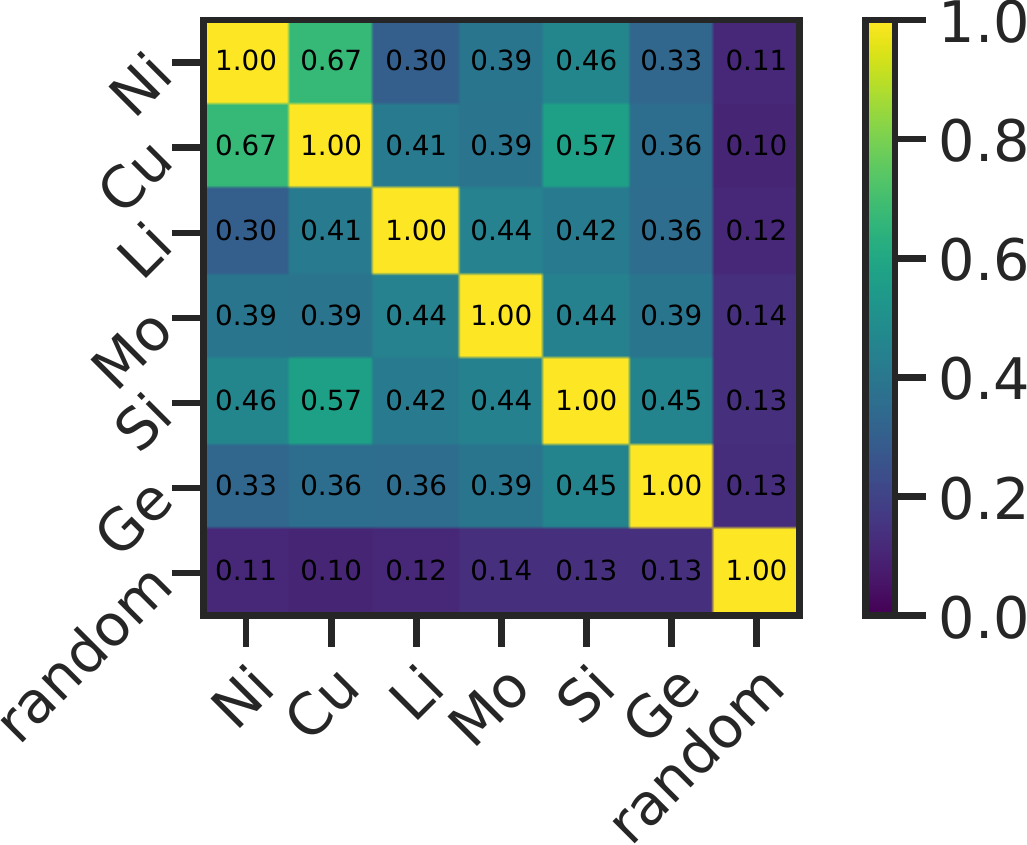}
    \caption{Heatmap displaying the fraction of components selected via LASSO regression that are shared between two materials for the case of SC radial basis functions.
    The penalty is term set to $5\cdot10^{-4}$.
    The random label indicates a set of 65 randomly selected components out of the available 360.}
    \label{fig:sm_lasso_shared_components_sc}
\end{figure}

\clearpage
\subsection{PCA and LASSO dimensionality reduction for NSC basis functions}
Here we report graphs that mirror Figures 5, 6, 7, and 8 of the main text, but for the case of NSC radial basis functions.
\begin{figure}[h!]
    \includegraphics[width=7.5cm]{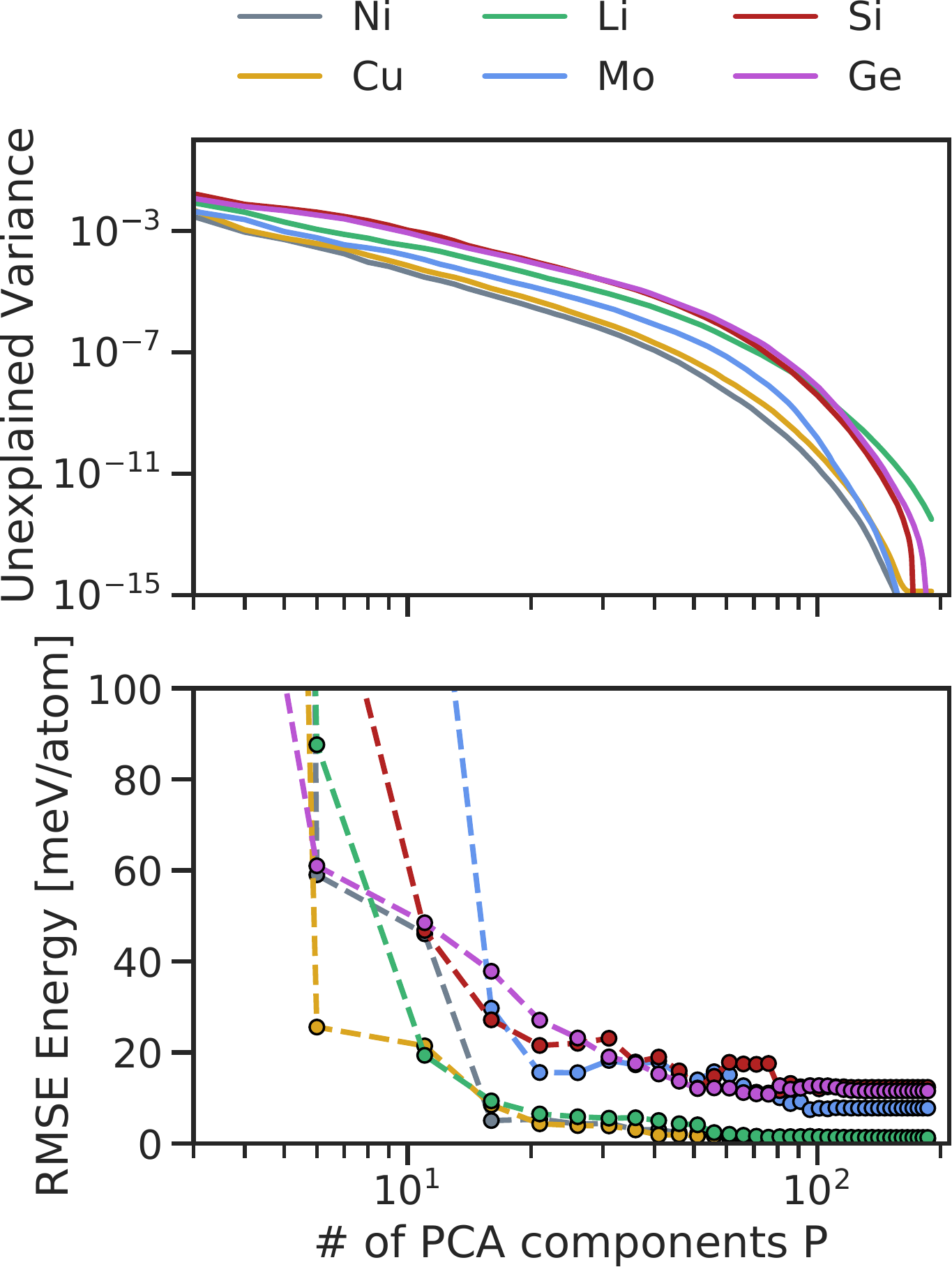}
     \caption{Top panel: data unexplained variance as a function of the number of PCA components accounted for.
     Bottom panel: RMSE on energies incurred by RR potentials employing the reduced descriptor $\mathbf{Q}^{PCA}_{P}$ on the validation set, as a function of the number of PCA components $P$.}
     %
    %  The colored crosses on the right indicate the RMSEs on energies incurred by a RR potential employing the full 289-dimensional descriptor $\mathbf{Q}$.}
    \label{fig:pca_nsc}
\end{figure}
\begin{figure}[h!]
    \includegraphics[width=6.5cm]{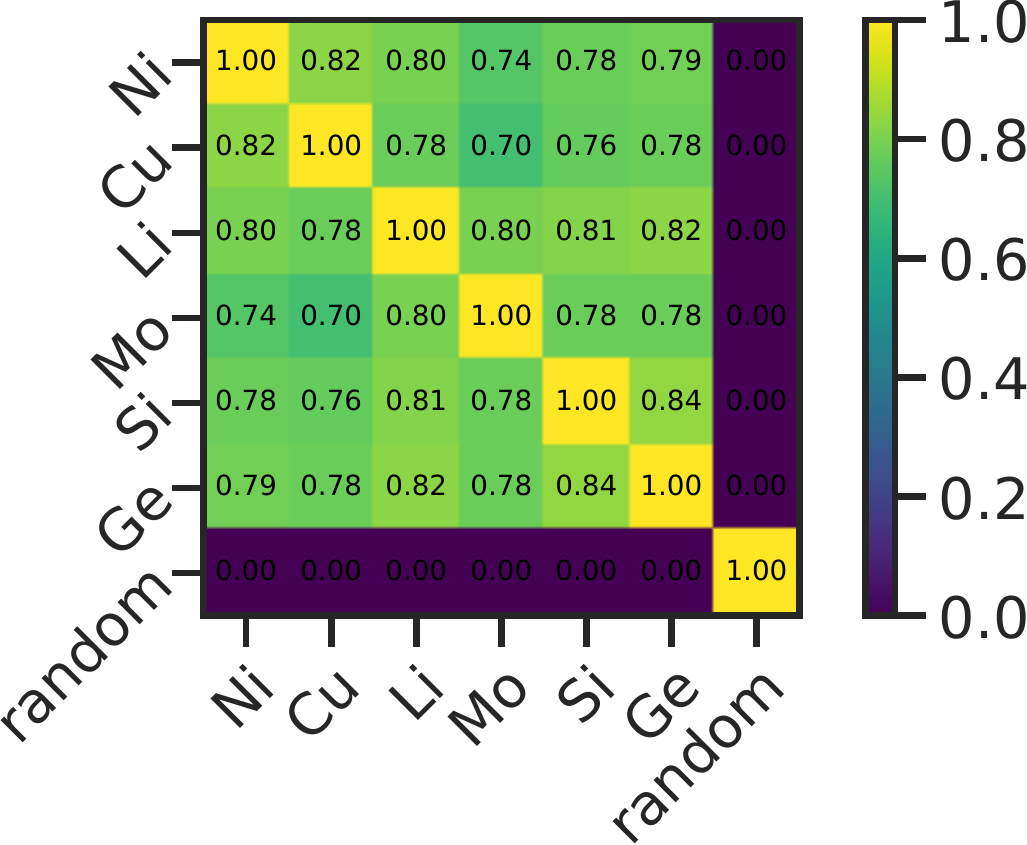}
    \caption{Heatmap displaying the fraction of dimensions shared by the sub-spaces generated by the first 80 PCA-selected directions of the descriptors $\mathbf{Q}$ among couples of single-element datasets.
    The random label indicates a sub-space generated by taking 80 random orthogonal vectors in the space of the $\mathbf{Q}$ vectors.}
    \label{fig:pca_shared_dimensions_nsc}
\end{figure}
\begin{figure}[h!]
    \includegraphics[width=7.5cm]{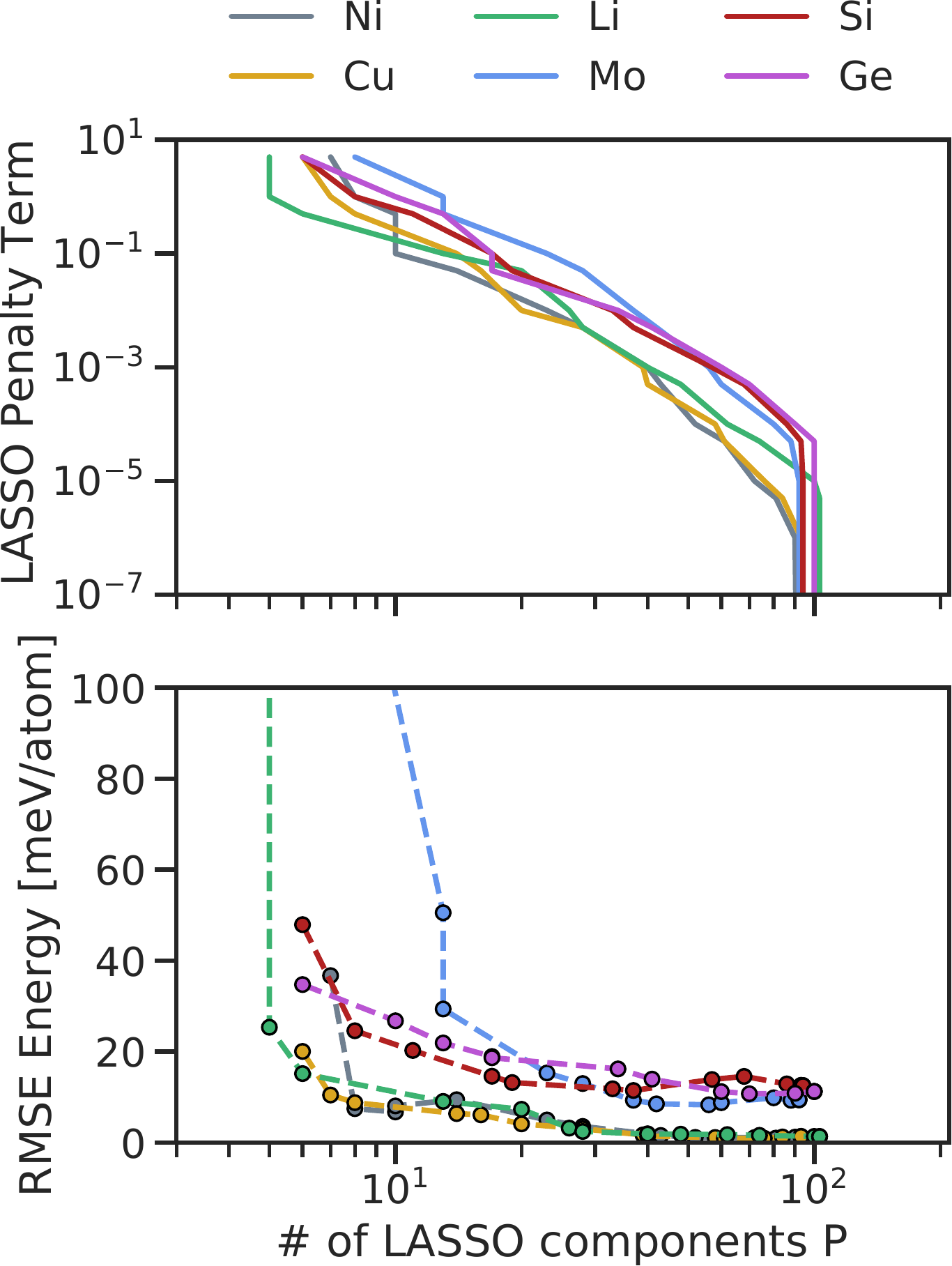}
     \caption{Top panel: LASSO penalty term as a function of the number of LASSO components accounted for.
     Bottom panel: RMSE on energies incurred by RR potentials employing the reduced descriptor $\mathbf{Q}^{LASSO}_{P}$ on the validation set, as a function of the number of LASSO components $P$.}
    \label{fig:lasso_nsc}
\end{figure}
\begin{figure}[h!]
    \includegraphics[width=6.5cm]{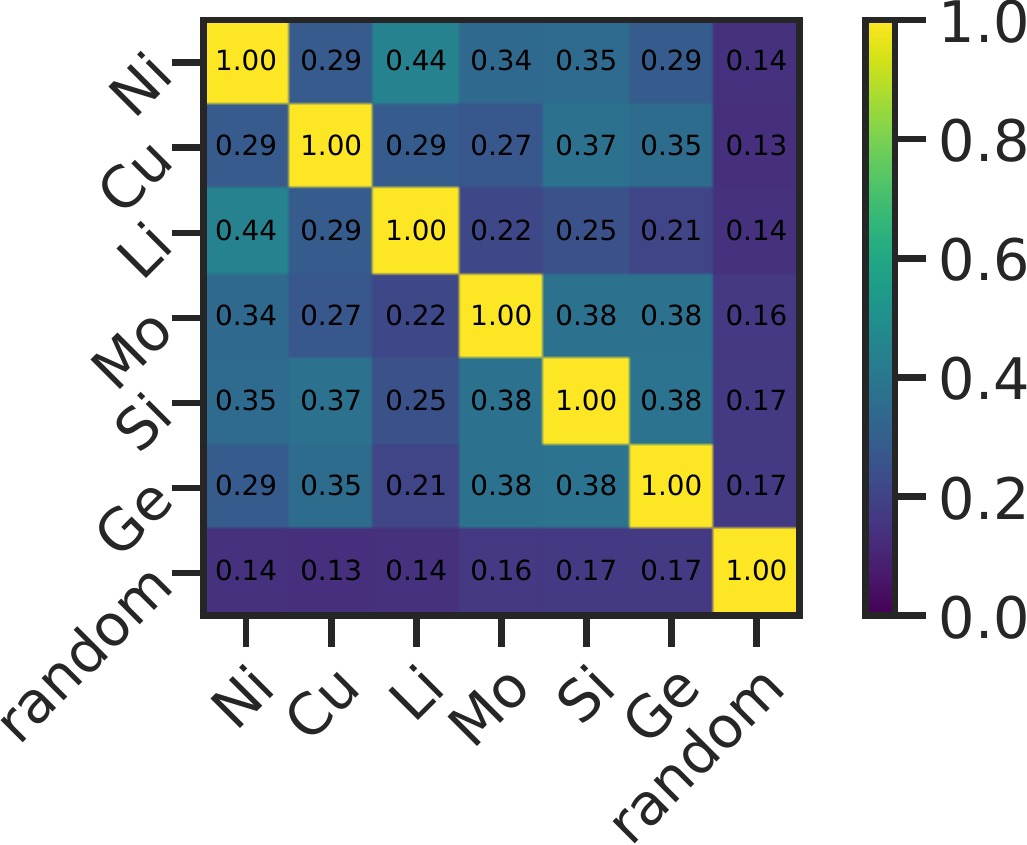}
    \caption{Heatmap displaying the fraction of components selected via LASSO regression that are shared between two materials for the case of SC radial basis functions.
    The penalty is term set to $5\cdot10^{-4}$.
    The random label indicates a set of 65 randomly selected components out of the available 360.}
    \label{fig:sm_lasso_shared_components_nsc}
\end{figure}

\clearpage
 \subsection{Dimension of the non-vanishing intersection of two sub-spaces}
To compute the dimension of the intersection $U \cap V$ of two sub-spaces $U$ and $V$ defined by the matrices of column basis vectors $\mathbf{U}$ and $\mathbf{V}$, we procede as follows.
If $\mathbf{z}$ is a vector that lies in $U \cap V$, then projecting $\mathbf{z}$ onto $U$ and $V$ does not change it:
\begin{equation}
\mathbf{z} = \mathbf{P}_{U}\mathbf{z} ~~~ \text{and} ~~~ \mathbf{z} = \mathbf{P}_{V}\mathbf{z},
\label{eq:intersection}
\end{equation}
where the projection matrix $\mathbf{P}_{U}$ is defined as \cite{basilevsky2013applied}:
\begin{equation}
\mathbf{P}_{U} = \mathbf{U} \left( \mathbf{U}^T \mathbf{U} \right)^{-1} \mathbf{U}^T.
\label{eq:projection_matrix}
\end{equation}
Therefore, we can write:
\begin{equation}
(\mathbf{P}_{U}\mathbf{P}_{V})\mathbf{z} =  \mathbb{I}\mathbf{z}.
\label{eq:intersection_2}
\end{equation}
We then define $\mathbf{M} = \mathbf{P}_{U}\mathbf{P}_{V}$ and calculate the dimension of the intersection of U and V as the number of eigenvalues of $\mathbf{M}$ that are close to 1 in value.
This calculation rigorously defines the dimension of the intersection of two sub-spaces when only eigenvalues equal to 1 are accounted for.
In our case study, we however find that the estimation of the dimension is truly informative only if we set a numerical threshold for counting the eigenvalues ``close'' to 1.
This is done to allow for ``almost matching'' dimensions to be also accounted for.
\begin{figure}[h!]
    \centering
    \includegraphics[width=8cm]{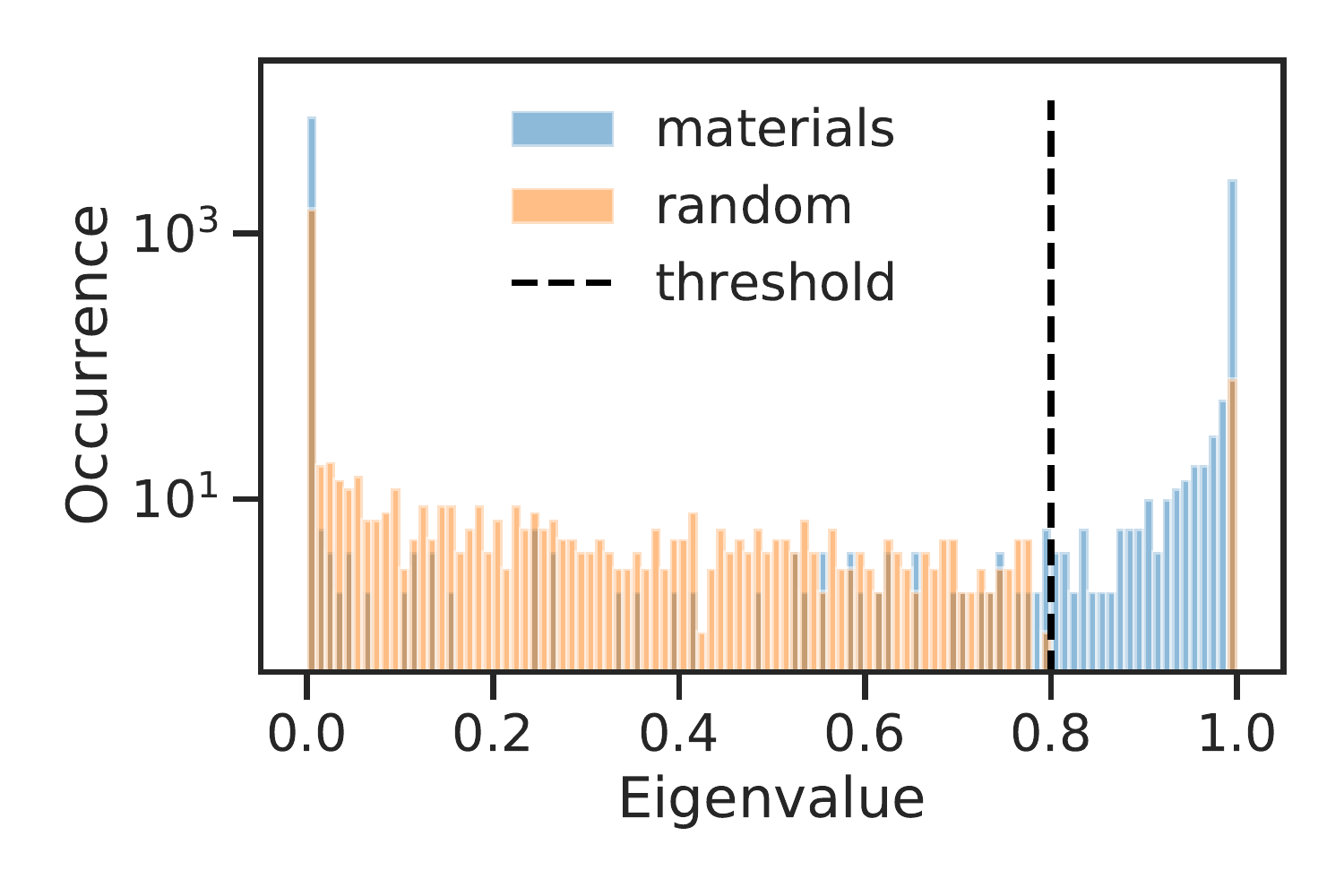}
    \caption{Occurrence of eigenvalues of the matrix $\mathbf{M}$ computed among PCA rotation matrices of materials descriptors (blue), and between PCA rotation matrices of materials descriptors and matrices containing 80 360-dimensional orthonormal vectors chosen at random (orange).
    The black dashed line indicates the value of the largest eigenvalue $< 1$ of $\mathbf{M}$ for the ``random'' case. ``random'' matrix $\mathbf{M}$.}
    \label{fig:pca_eigenvalues}
\end{figure}

To set the threshold value, we look at the eigenvalue distribution for $\mathbf{M}$ in two cases: when U and V are both defined by the first 80 components of the PCA rotation matrices for any two single-element datasets, and when U is instead defined by 80 360-dimensional orthonormal vectors chosen randomly, and V is defined as in the previous case.
In Fig.\ref{fig:pca_eigenvalues} we see the eigenvalue distribution for the two cases in blue and orange, respectively.
We choose the threshold to be the value of the largest eigenvalue $< 1$ of $\mathbf{M}$ for the ``random'' case.
For Fig.~5 of the main text, the threshold value is set to 0.8.

\end{document}